\documentclass[pdftex,twocolumn,epjc3]{svjour3}          

\RequirePackage[T1]{fontenc}

\smartqed  
\RequirePackage{lineno}

\RequirePackage{graphicx}
\RequirePackage{mathptmx}      
\RequirePackage{flushend}
\RequirePackage[numbers,sort&compress]{natbib}
\RequirePackage[colorlinks,citecolor=blue,urlcolor=blue,linkcolor=blue]{hyperref}
\usepackage{graphicx}	
\usepackage{amsmath}	
\journalname{Eur. Phys. J. C}

\newcommand{\Scc}[1]{{\color{black}{{ #1}}}}
\newcommand{\Comm}[1]{{\color{black}{{ #1}}}}

\begin{document}

\title{Analyzing the speed of sound in neutron star with machine learning}


\author{Sagnik Chatterjee\thanksref{e1,addr1}
        \and
        Harsha Sudhakaran\thanksref{e2,addr1} 
         \and
        Ritam Mallick\thanksref{e3,addr1}
}

\thankstext{e1}{e-mail: sagnik18@iiserb.ac.in}
\thankstext{e2}{e-mail: sudhakaran.harsha@gmail.com}
\thankstext{e3}{e-mail: mallick@iiserb.ac.in}

\institute{Department of Physics, Indian Institute of Science Education and Research Bhopal, Bhopal, Madhya Pradesh, 462066, India \label{addr1}}

\date{Received: date / Accepted: date}

\maketitle

\begin{abstract}
Matter properties at the intermediate densities are still unknown to us. In this work, we use a neural network approach to study matter at intermediate densities to analyze the variation of the speed of sound and the measure of trace anomaly considering astrophysical constraints of mass-radius measurement of 18 neutron stars. Our numerical results show that there is a sharp rise in the speed of sound just beyond the saturation energy density. It attains a peak around $3-4$ times the saturation energy density and, after that, decreases. \Scc{This hints towards the appearance of new degrees of freedom and smooth transition from hadronic matter in massive stars. The trace anomaly is maximum at low density (surface of the stars) and decreases as we reach high density. It approaches zero and can even be slightly negative at the centre of massive stars. It has a negative trough beyond the maximal central densities of neutron stars. The change in sign of the trace anomaly hints towards a near-conformal matter at the centre of neutron stars, which may not necessarily be conformal quark matter.}
\end{abstract}

\section{Introduction} \label{sec:intro}

The matter at neutron star (NS) cores is in a highly compressed state, and due to gravity, density can build up to a few times nuclear saturation density. They are very compact and have been observationally identified as pulsars, with their mass ranging from $0.7 - 2.1$ solar masses ($M_{\odot}$) \citep{Valentim} and radius between $10-15$ km. They are, therefore, one of the best laboratories to test the theory of strong interaction at the high-density low-temperature regime. One of the most interesting facts of QCD is the conjecture of a phase transition (PT) \citep{Witten} from hadronic matter (HM) to quark matter (QM) at these densities \citep{Shuryak}. This has led to extensive experimental/observational and theoretical efforts to understand the matter properties at neutron star interiors. The field got a considerable boost after several precise mass measurements of massive pulsars \citep{Riley_2019, Riley_2021,Miller_2019, Miller} and the gravitational wave (GW) observation of binary NS mergers (BNSM) \citep{Abbot}. Although we have been able to constrain the matter properties (equation of state (EoS)) in the last few years, our knowledge about the EoS at such densities is limited.

The first principle calculation of matter at low density (till saturation density) \citep{Hebeler_2013} and at asymptotic high density are possible \citep{Lattimer}. However, intermediate densities are still beyond our reach. Physicists have either relied on models or used agnostic approaches to understand the matter at intermediate densities found at the cores of NSs \citep{deForcrand}. The agnostic approach fails to have any microscopic detail of the matter; however, the overall properties of the NS can be obtained from them. Whatever the approach, the models must satisfy the constraints obtained from theory and observations. EoS obtained from the agnostic approach are based on methods like Bayesian analysis and machine learning to predict the most probable EoS. Bayesian analysis is extensively used nowadays to impose constraints from observational data like in BNSM \citep{Most, Annala, Zhang_2018}. They are statistical or probabilistic models that account for uncertainty in measurements.

With the current constraints from observational astrophysics of neutron stars and from the agnostic models, one can restrict the EoS to a certain limit. To constrain the EoS, one needs a good parameter to build the agnostic model. One of the parameters is the adiabatic speed of sound ($c_{s}$), which relates the density and pressure of the EoS \citep{Bedaque}. It is defined as the first derivative of pressure with respect to energy density; and refers to the amount of matter that can be accommodated for a particular pressure so that it can balance the gravitational pull. This is the measure of the stiffness of the EoS. The sound speed is constrained between $0$-$1$, resulting from thermodynamic stability and causality. Large sound speed refers to greater stiffness and, thereby, massive stars. The sound speed is also bounded at asymptotic densities by perturbative-QCD (pQCD) calculations, which predict that the sound speed square should asymptotically reach the value $1/3$ \citep{Komoltsev}. However, it should be emphasized that in the intermediate densities, the sound speed could breach this value, and it has been shown that it should be necessary to generate massive stars \citep{Kojo2021,Altiparmak_2022}.

There are many formulations by which one can optimize the EoS using the constraints on the speed of sound. Machine learning (ML) can be used to study various nuclear and matter properties of NS \citep{Fujimoto2018,Fujimoto_2020,Fujimoto2021,Morawski, Ferreira_2021,Ferreira,Krastev, Traversi_2020, Soma_2022}. The optimal speed of sound can be obtained by using the ML approach with the observational constraints and mass-radius measurements of pulsars. Supervised ML can be used to map the non-linear mass-radius relation to the speed of sound. Non-linear mapping is done by the neural network (NN), and the mass-radius uncertainty is used to learn the map and then predict the sound speed. The data are prepared in a model-independent manner from observations and then fed into the neural network. Having a good amount of data can minimize the error in predicting the correct speed of sound and can be independent of any unintended bias. In this work, we randomly build agnostic EoS from the sound speed, employ NS mass constraints, and prepare our data from several mass-radius measurements of observed pulsars. The mass-radius data is used as an input to find the speed of sound as an output from the NN. The speed of sound is then used to obtain the optimal matter properties at NS cores.

Although most of the work using ML in NSs has been concerned mainly with constraining matter properties close to the nuclear saturation density, in this work, we try to use NN to delve deeper into matter properties at intermediate densities. This work uses data from astrophysical observations to construct an NN structure. Using a new set of EoS without prior information about the astrophysical constraints, we reconstruct a new set of EoSs using the NN. Using these NN informed EoS, statistical analysis of the variation of the speed of sound and trace anomaly is performed. Our analysis shows that the speed of sound has a rapid rise beyond twice the saturation energy densities and reaches a maximum around three times the saturation density. Therefore, the matter properties inside $1.6 M_{\odot}$ and $2.2 M_{\odot}$ stars are significantly different.

The paper is arranged in the following way: Section \ref{sec:Formalism} gives the formalism of our work, which explains the data preparation and the formulation of the NN. Next, we show our results in section \ref{sec: Results}. Finally, section \ref{sec:Summary} summarizes the work and draws essential conclusions.

\section{Formalism} \label{sec:Formalism}

In this article, our main aim is to predict the optimal speed of sound inside the neutron star using ML. One of the main bottlenecks of using ML is the need for a massive amount of data for reliable results. We construct a huge number of agnostic EoSs by randomizing the speed of sound in the density range where it is unknown. Then, implementing several mass radius observations of pulsars, we sample our data. We introduce errors in the mass-radius measurement from observations. The data generation starts with constructing an enormous amount of agnostic EoS, satisfying the thermodynamic stability criterion and respecting the condition that the speed of sound is bounded in the range $0 \le c_s \le 1$. The EoS at the low-density and high-density regimes is simultaneously constrained with BPS EoS and pQCD calculation.

\subsection{Data Preparation} \label{sec:Data}
Generating EoS using an agnostic approach can be done in three ways. The first technique uses a polytropic EoS \citep{Raithel_2016, Raithel_2022}. Second, one uses the spectral decomposition technique to construct EoSs \citep{Lindblom_2010}. The third approach uses the randomization of the speed of sound, which we adopt in the present work for constructing our EoS. As $c_s$ can vary from $0-1$, it allows us to scan the entire EoS space (the pressure and energy density space), thus allowing us to perform a detailed analysis.

To begin with, we randomly generate an ensemble of agnostic EoSs. This is done by taking BPS crust EoS \citep{Baym1971} till a density of $n \leq 0.57 n_{0}$ ($n_{0} = 0.16 fm^{-3}$  being the nuclear saturation density) . From this density to a density $1.1 n_{0}$, we take either "stiff" or "soft"  EoS \citep{Hebeler_2013, Keller, Epelbaum2009, Tews} spanning the chiral effective theory (CET) band. 
Corresponding to the saturation density, the saturation energy density is defined as $\epsilon_0$. From $\epsilon_{0}$ to $10 \epsilon_{0}$, we randomize the square of the speed of sound ($c_{s}^2$) with six equal spacing assuming that the maximum density that a star can attain (for any EoS) can at most be $10$ times saturation density. \Scc{This is sufficient for analyzing NSs; however, QCD predicts that conformal symmetry is restored at asymptotically high densities. Therefore, as a check, we extend the speed of sound randomization till $90 \epsilon_0$, where the speed of sound is $1/3$.}
We randomize the speed of sound at three points, which are $25 \epsilon_{0}$, $55 \epsilon_{0}$, and $90 \epsilon_{0}$. The $c_{s}$ value cannot exceed the causal limit or be less than zero. Hence, we take the range as $0 \leq c_{s}^{2} \leq 1$. As $c_{s}^{2} = (\partial P/\partial \epsilon)_{s}$, where $P$ is the pressure and $\epsilon$ is the energy density; hence we perform a linear interpolation at each of these intervals to obtain the entire profile of the EoSs. At asymptotically high densities, our EoSs follow the conformal field theory ($c_{s}^{2} \rightarrow 1/3$) \citep{Kurkela_2010,Gorda_2018,Altiparmak_2022}. 

\begin{figure}
	\centering
	\begin{minipage}[ht]{0.5\textwidth} 
		\includegraphics[scale=0.5]{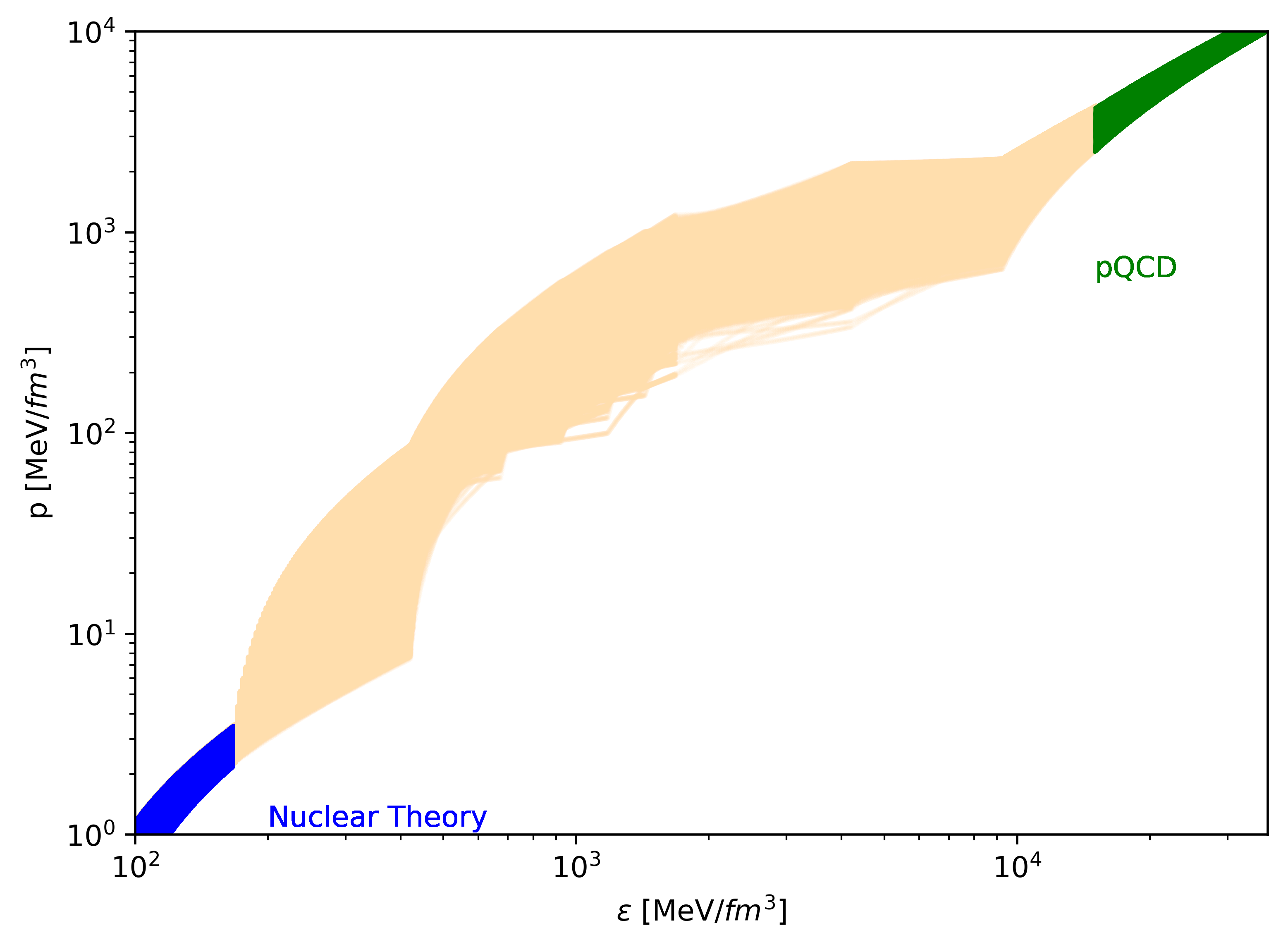}
	\end{minipage} 
	\begin{minipage}[ht]{0.5\textwidth} 
		\includegraphics[scale=0.58]{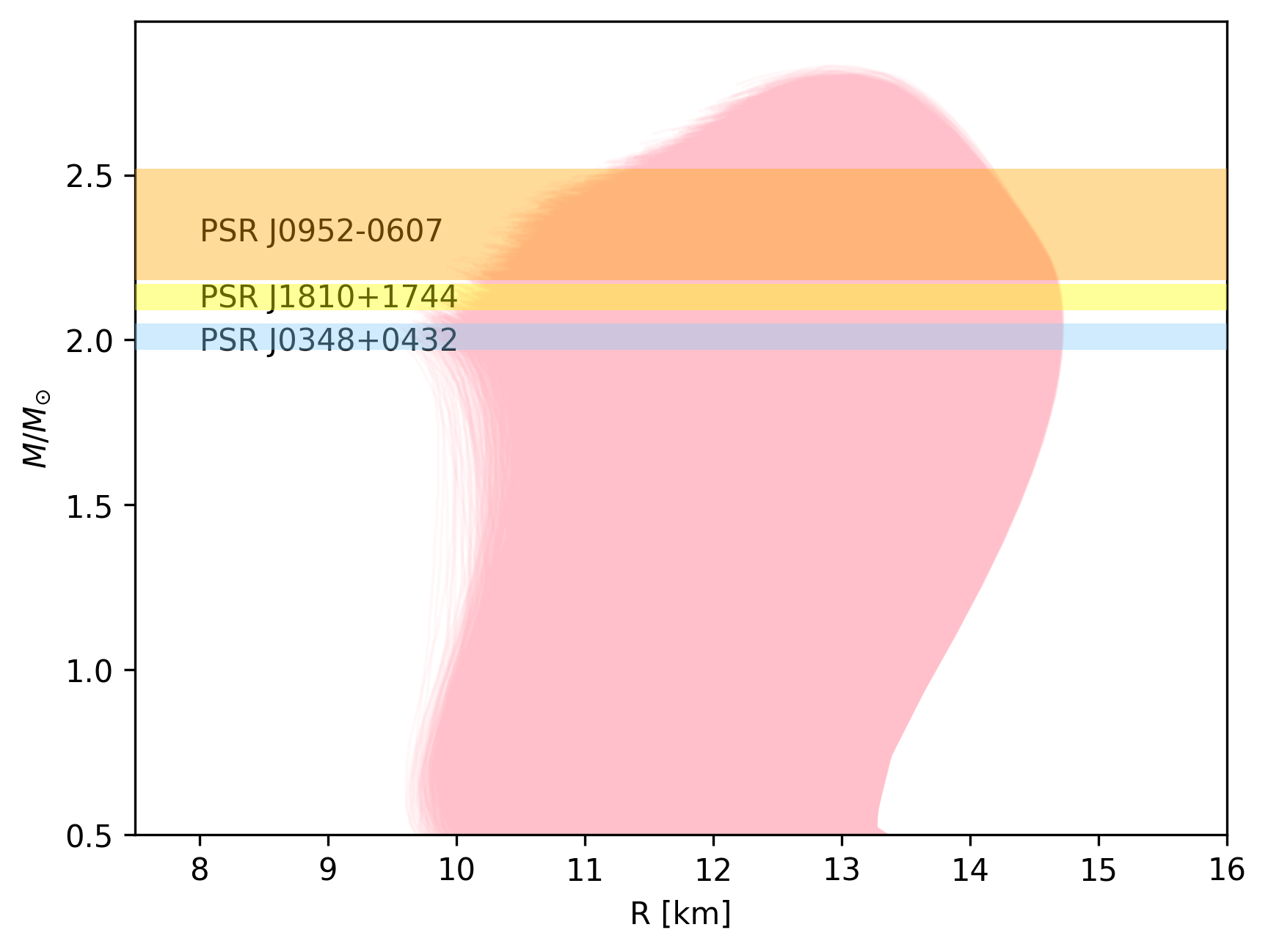}
	\end{minipage} 
	\label{eos}
\end{figure}

We created $40,000$ such EoSs as shown in Fig \ref{eos}. For our analysis, we only consider the EoSs which satisfy the maximum mass limit of $M_{TOV} \ge 2.0 M_{\odot}$. We need to solve the Tolman-Oppenheimer-Volkoff equation (TOV) \citep{TOV} and generate a mass-radius (M-R) sequence. From these M-R sequences, we reject those EoSs having maximum mass $M_{TOV} < 2 M_{\odot}$ imposed by the mass measurements of J0348+0432 \citep{Antoniadis} ($M = 2.01 \pm 0.04 M_{\odot}$) and J0740+6620 \citep{Cromartie} ($M = 2.08 \pm 0.07 M_{\odot}$). 

In this work, we have not imposed strict conditions on tidal deformability \citep{Hinderer} and the NICER constraints given by \citep{Riley_2019,Riley_2021,Miller, Miller_2019}. This is done to have enough EoS for training and testing our Neural network and to prevent any biases which might arise due to constraints of the EoS. 

\begin{figure*}[ht]
	\centering
	\includegraphics[scale=0.40]{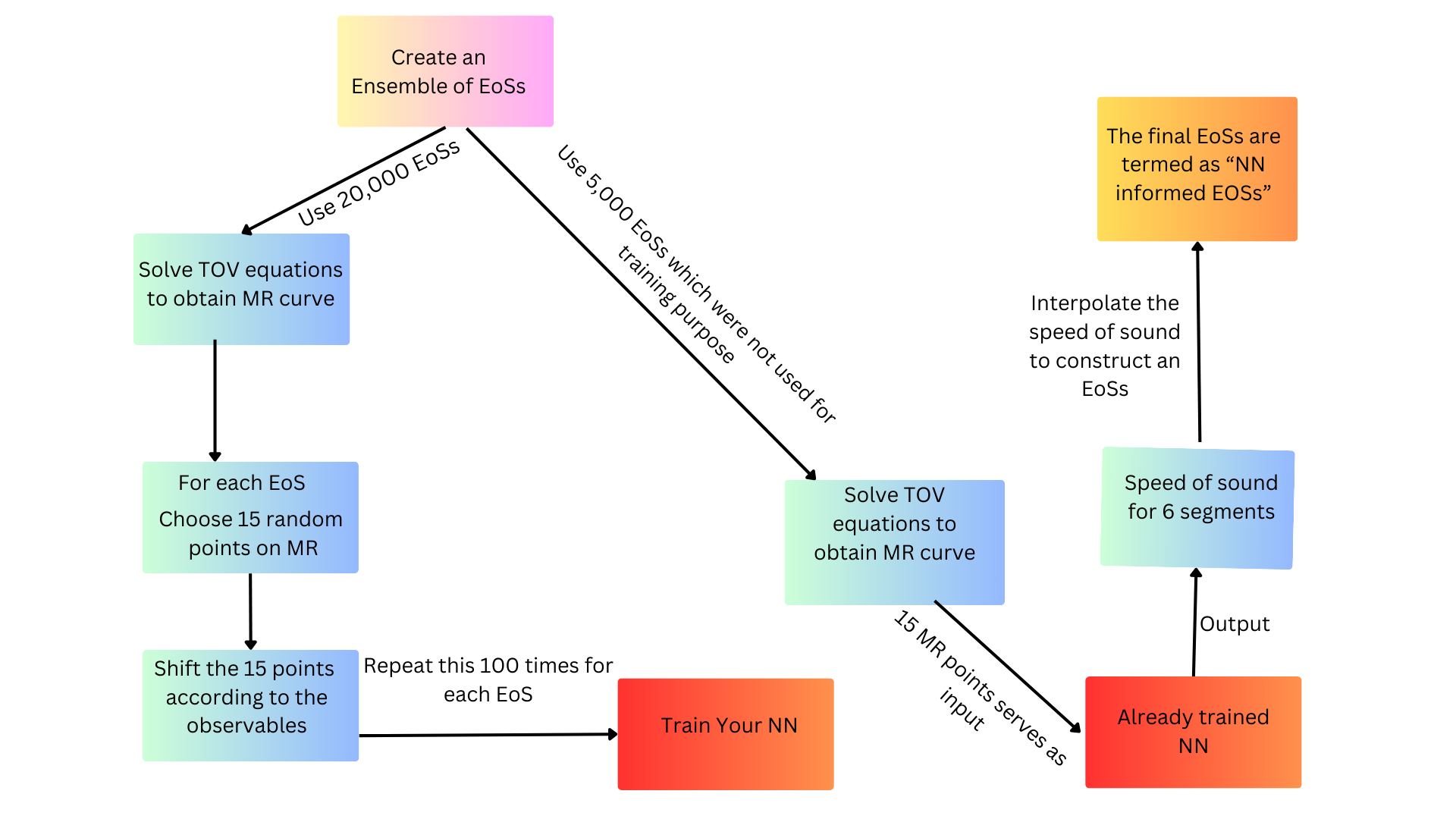}
	\caption{\small Flowchart showing the entire process of our analysis. }
	\label{steps}
\end{figure*}

Next, we segregate our EoSs into a training-set ($20,000$ EoSs) out of the $40,000$ EoSs (the entire Flowchart of our process can be found in fig \ref{steps}). Theoretically generated EoS has a smooth sequence stretching from very low to maximum mass. Also, there is no uncertainty in the mass and radius for a given EoS. However, observationally, the mass and radius measurements are accompanied by errors. To build the training data from the constructed EoS, we need to account for these errors in our data. We select $15$ points on the MR curve for each EoS to train our NN. We use $18$ pulsar mass radius data as shown in table \ref{obs:table} given by \cite{Soma_PRD}. The values are determined by marginalizing over the neutron star properties using a Gaussian distribution over various observations \citep{Ozel_2016, rNattila, Gonzalez, Riley_2019}.

\begin{table}
\centering
\caption{Observable Mass ($M_{\odot}$) and Radius (km)\citep{Soma_PRD} } \label{obs:table}
\begin{tabular}{|l|c|c|}
\hline
Object & Mass ($M_\odot$) & Radius (km) \\ \hline
M13 & $1.42 \pm 0.49$ & $11.71 \pm 2.48$ \\
M28 & $1.08 \pm 0.30$ & $8.89 \pm 1.16$ \\
M30 & $1.44 \pm 0.48$ & $12.04 \pm 2.30$ \\
NGC 6304 & $1.41 \pm 0.54$ & $11.75 \pm 3.47$ \\
NGC 6397 & $1.25 \pm 0.39$ & $11.48 \pm 1.73$ \\
$\omega$ Cen & $1.23 \pm 0.38$ & $9.80 \pm 1.76$ \\
4U 1608-52 & $1.60 \pm 0.31$ & $10.36 \pm 1.98$ \\
4U 1724-20 & $1.79 \pm 0.26$ & $11.47 \pm 1.53$ \\
4U 1820-30 & $1.76 \pm 0.26$ & $11.31 \pm 1.75$ \\
EXO 1745-24 & $1.59 \pm 0.24$ & $10.40 \pm 1.56$ \\
KS 1731-26 & $1.59 \pm 0.37$ & $10.44 \pm 2.17$ \\
SAX J1748.9-2021 & $1.70 \pm 0.30$ & $11.25 \pm 1.78$ \\
X5 & $1.18 \pm 0.37$ & $10.05 \pm 1.16$ \\
X7 & $1.37 \pm 0.37$ & $10.87 \pm 1.24$ \\
4U 1702-42 & $1.90 \pm 0.30$ & $12.40 \pm 0.40$ \\
PSR J0437–4715 & $1.44 \pm 0.07$ & $13.60 \pm 0.85$ \\
PSR J0030+0451 & $1.44 \pm 0.15$ & $13.02 \pm 1.15$ \\
PSR J0740+6620 & $2.08 \pm 0.07$ & $13.70 \pm 2.05$ \\ \hline
\end{tabular}
\end{table}

\begin{figure}[ht]
	\centering
	\includegraphics[scale=0.50]{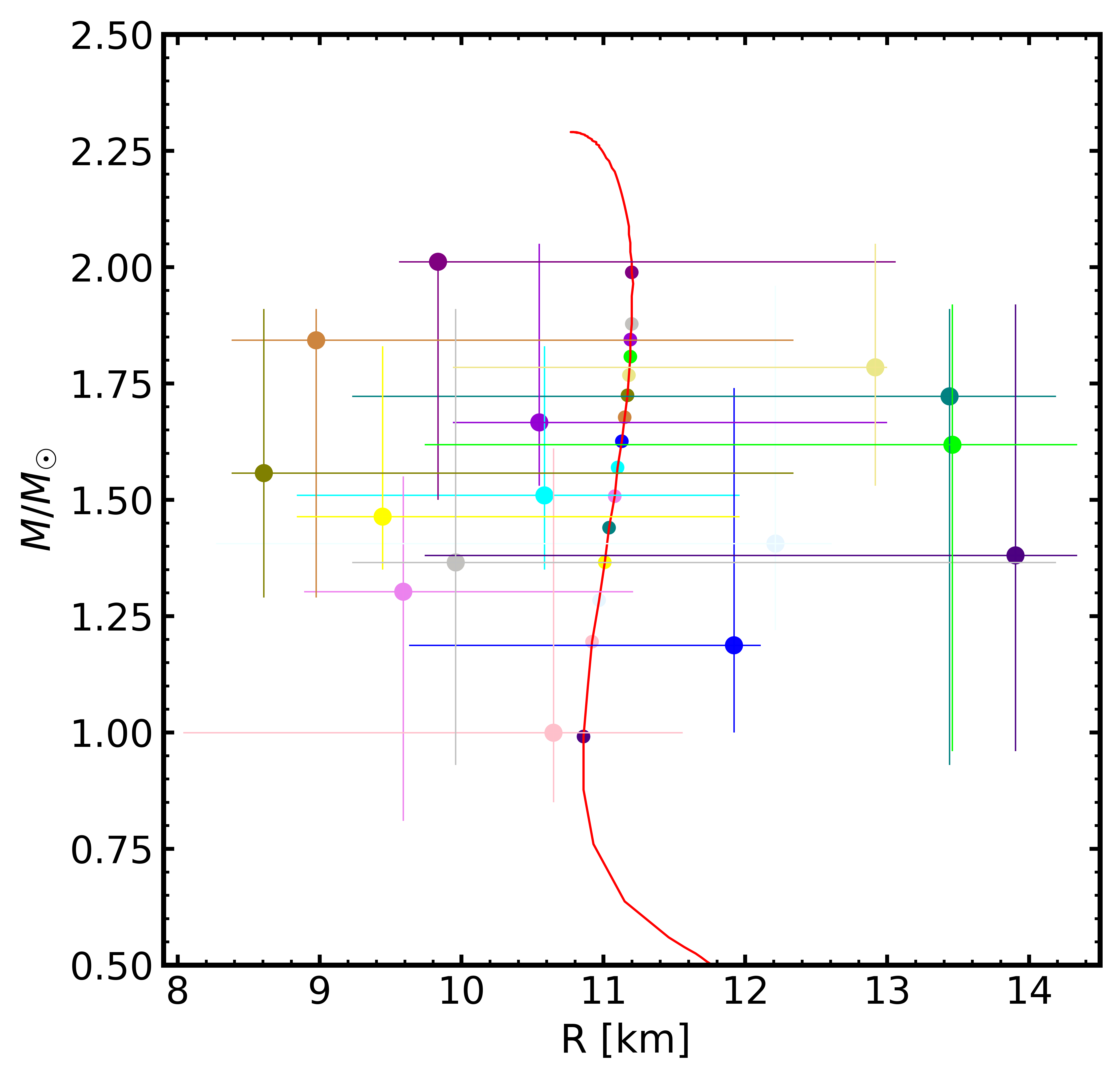}
	\caption{\small The following figure shows how we sample 15 random points on the MR curve of our EoS and then randomly sample and shift each of those 15 points. The red coloured line denotes the MR curve, and 15 points that are randomly chosen on the MR curve are shown in different colours. Each point is shifted according to the observables defined in table \ref{obs:table}. The shifted point corresponding to a single point on the MR is shown in the same colour, with the observation error also denoted by the same colour.}
	\label{error}
\end{figure}

We check under which observation range (table \ref{obs:table}) our randomly chosen point on the MR curve lies and identify it. We then shift it randomly but within the error range of that given observation provided in table \ref{obs:table}. For example, in fig \ref{error}, consider the first point (purple colour with a mass of $1.98 M_{\odot}$), which has been randomly chosen on our MR curve. Next, we check under which observations the mass $1.98 M_{\odot}$ lies. The particular mass point lies within the range of the observation of "4U 1724-20", "4U 1724-20", "SAX J1748.9-2021", "4U 1724-42", and "PSR J04740+6620". Out of these five observations, we again randomly choose one observation (which is "SAX J1748.9-2021" for this particular example) and then shift its mass and radius according to the variance of that particular observation from table \ref{obs:table}. 
For each EoS, this process was repeated 50 times, each time randomly choosing a new set of 15 points on the MR curve. We have $20,000$ EoSs, each of which is iterated $100$ times. Hence we have ($20,000 \times 100$) of $20,00,000$ sets of data, each having $15$ data points (representing the $15$ points on the MR curve).
\subsection{Neural Network} \label{sec:NN}

The measurement of the maximum mass, along with NICER and LIGO observation of NS, has severely constrained the EoS; however, the most likely EoS for high-density matter, mainly at the intermediate region, still needs to be determined with relative confidence. In the last few years, there has been a continuous effort to extract the most likely EoS, and one of the most used methods to determine them is using Bayesian Analysis \citep{Raithel,Lim2019,Char,Biswas_2022}. There have also been methods employing ML using NN for determining the likely EoS \citep{Fujimoto2018, Morawski}. In this article, we employ ML methods using NN \citep{Goodfellow-et-al-2016}; however, we use the speed of sound inverse mapping to find the EoS. We use the mass-radius data (obtained from the previous subsection) as an input to the neural network to obtain the $c_s^{2}$ for the six segments ($\epsilon$). From the $c_s^{2}$ obtained from the NN, we find the corresponding pressure and regenerate the ML predicted EoS.  

\begin{figure}
	\centering
	\includegraphics[scale=0.37]{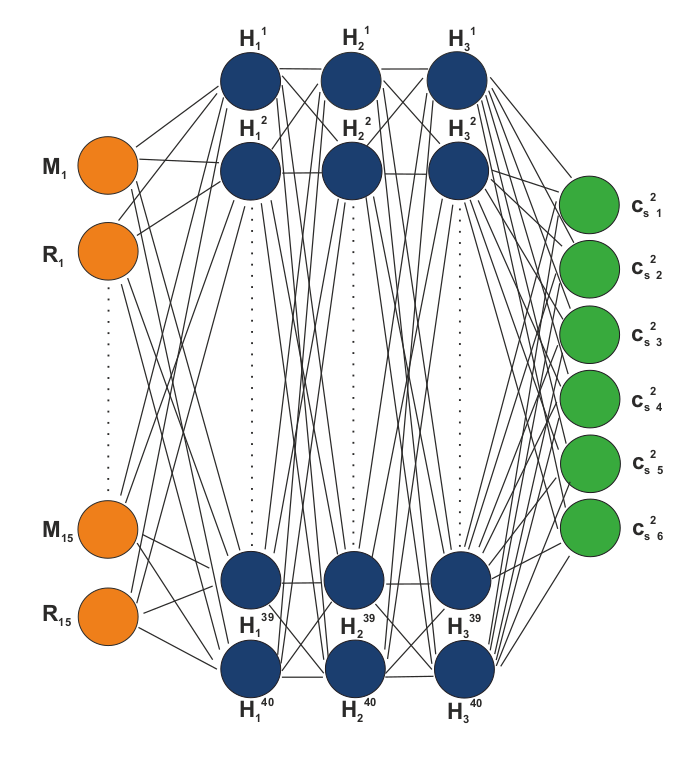}
	\caption{\small A schematic diagram of the structure of our NN. The input layers are shown with orange colours, with $M_{1}, M_{2}...M_{15}$ and $R_{1}, R_{2},..., R_{15}$ are the $15$ pairs of mass and radius which we provide as input for our NN. The hidden layer nodes are shown in blue, and each hidden layer is marked as $'H'$. The final output layers are represented by green colour having six $c_{s}^{2}$ as the outputs.} 
	\label{NN}
\end{figure}

\begin{figure}
	\centering
		\includegraphics[scale=0.60]{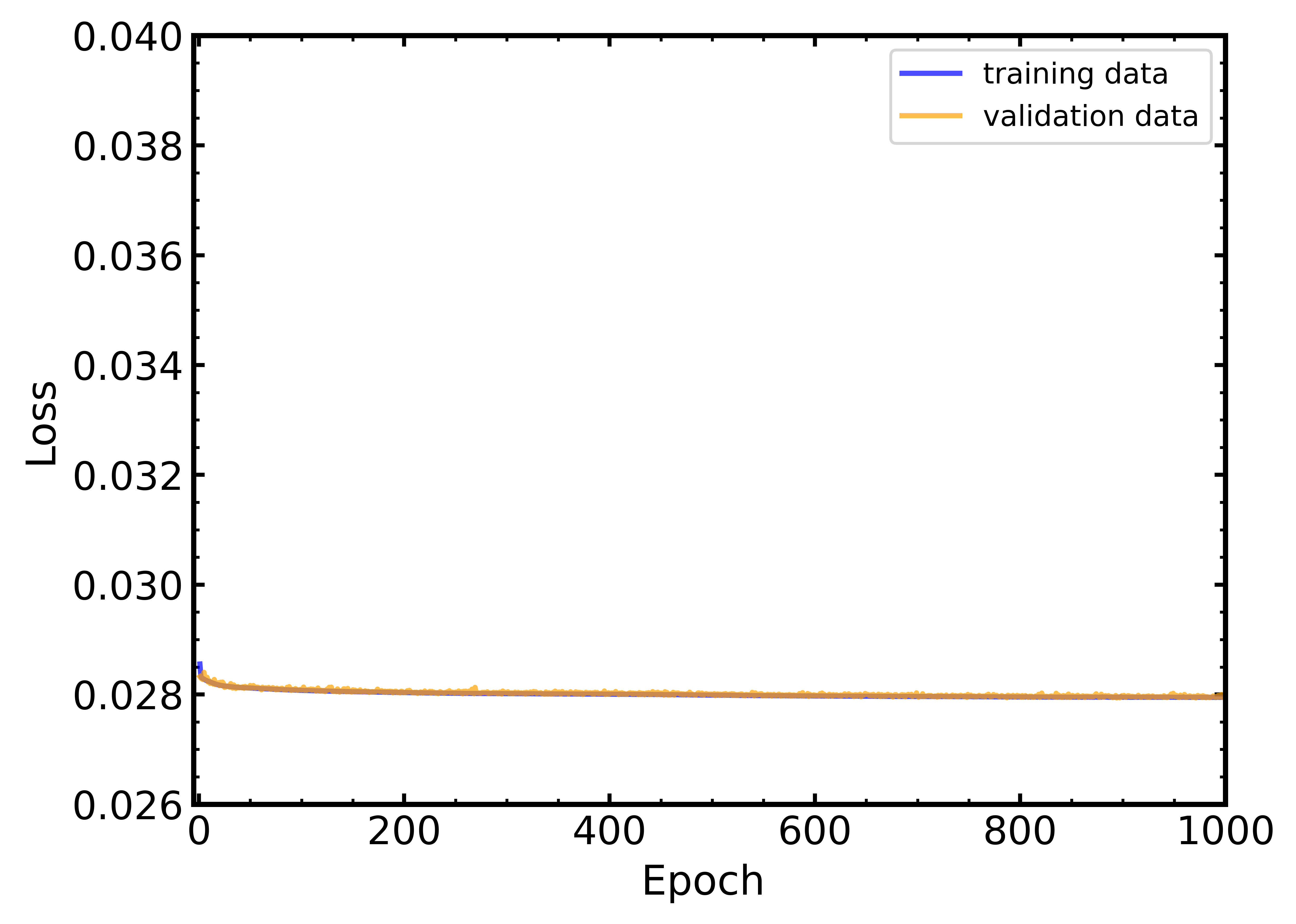}
	\caption{Plot for Loss (mean squared logarithmic error) against epochs. The blue-coloured line represents the change in mean squared logarithmic error for training data. In contrast, the orange-coloured line represents the change in mean logarithmic squared error for validation data up to 1000 epochs. }
	\label{errorvsepoch50}
\end{figure}

The NN structure has an input layer with $30$ neurons that take $15$ randomly sampled pairs of M and R as input. Subsequently, we have three hidden layers, each having $40$ neurons. The final layer has $6$ neurons which give us the $c_{s}^{2}$ as output as shown in fig \ref{NN} and represented in table \ref{tab:my_label}. Since the maximum central density of the NSs lies in the range of $\epsilon_{0}$ to $10 \epsilon_{0}$, we take the output neuron to be 6, which is equivalent to the number of segments made while constructing our EoSs. The data obtained after sampling is split into training and validation data. About $80$ percent is the training data, and the rest $20$ percent is the validation data. The model learns the relevant parameters from the training data, and its performance is evaluated on the validation data. The data points are normalized with $3 M_{\odot}$ for the mass and $20 \: \text{km}$ for the radius. This normalization speeds up the learning procedure, and as a result, the loss function converges faster. Also, training data is prepared with data points having mass greater than $0.5 M_{\odot}$. The loss function is the difference between actual output and predicted output. For our model, we take MSE (mean squared logarithmic error) as our loss function, which is optimized using \Scc{Nestrov-Adam or Nadam \citep{dozat2016}. Nadam uses momentum and adaptive learning rates to make optimization more efficient. The activation function for hidden layers is Mish, a self-regularized non-monotonic function. Nadam and Mish were chosen after a robust analysis of different optimizers and activation functions as shown in table \ref{table_opt} of \ref{appx:opt}}. For the output layer, the activation function is \textit{tanh}. \textit{tanh} returns values in the range from 0 to 1 for positive inputs and, therefore, can be used for predicting $c_{s}^{2}$, which also lies in the same range. Our model converges to a training loss of 0.0279, as shown in fig \ref{errorvsepoch50}. Our model uses the Python Keras \citep{chollet} with TensorFlow \citep{tensorflow}. Our results were performed over different layer sizes with different learning rates. The results documented here are the optimal ones giving the minimum value of MSE. 

\begin{table}
	\centering
	\begin{tabular}{ | l| l| l| } 
		\hline
		\textbf{Layer} & \textbf{Neurons} & \textbf{Activation} \\ 
		\hline
		Input Layer & 30 & None \\ 
		\hline
		Hidden Layer 1 & 40 & Mish\\  
		\hline
		Hidden Layer 2 & 40 & Mish\\  
		\hline
		Hidden Layer 3 & 40 & Mish\\  
		\hline
		Output Layer & 6 & $\tanh$\\
		\hline
	\end{tabular}
	\caption{This table explains the structure of our neural network. The input layer contains 30 neurons. The subsequent three hidden layers each contain $40$ neurons with activation as Mish. The output layer has $6$ neurons with \textit{tanh} as the activation function}
	\label{tab:my_label}
\end{table}

\section{Results} \label{sec: Results}

\begin{figure}
	\centering
	\includegraphics[scale=0.55]{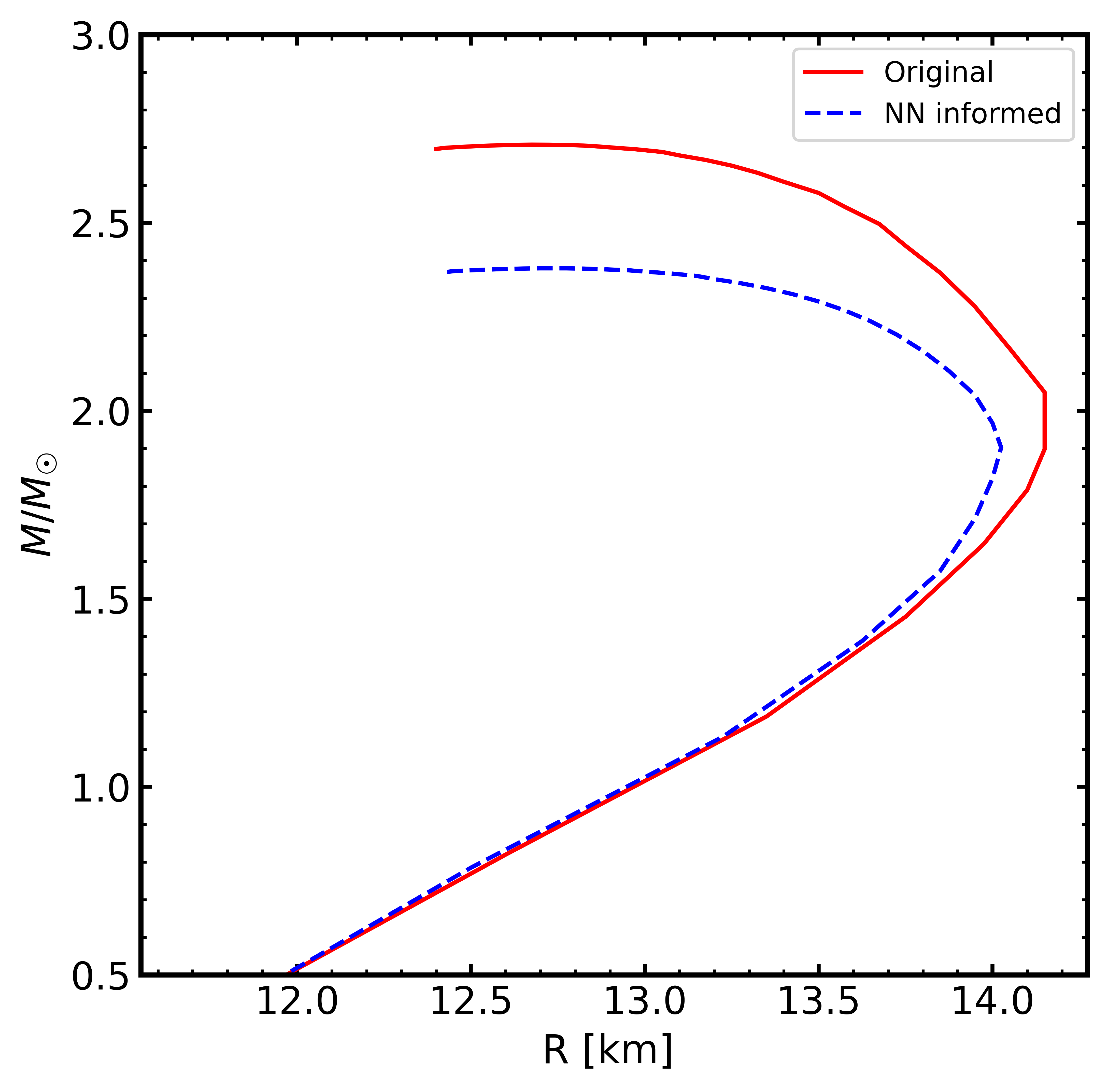}
	\caption{\small The red coloured curve represents an initial MR curve corresponding to a sample EoS. Fifteen random points on this MR curve are chosen and fed into our NN, which then gives the speed of sound for each of our six segments as output. These speeds of sounds were then used to interpolate our energy densities, and we reconstructed an EoS. The blue-coloured MR curve represents the MR curve of such a NN informed EoS.}
	\label{MR_NN}
\end{figure}

Once we have described our formalism of constructing the agnostic EoS and building the data, we feed the data to the neural network to obtain the speed of sound for $6$ points within the star. The other $3$ points at much larger densities are obtained by interpolating them. From the speed of sound and for a given segment (in $\epsilon$), we interpolate them linearly to obtain the NN informed EoS. The NN informed EoS is then solved to obtain a mass-radius curve. 
Fig \ref{MR_NN} shows the nature of a NN informed MR curve from our trained NN. The red line indicates an original MR; this EoS serves as input to the NN to obtain a NN informed EoS. The original and NN informed MR curves are shown in red and blue, respectively.

\begin{figure}
	\centering
	\includegraphics[scale=0.61]{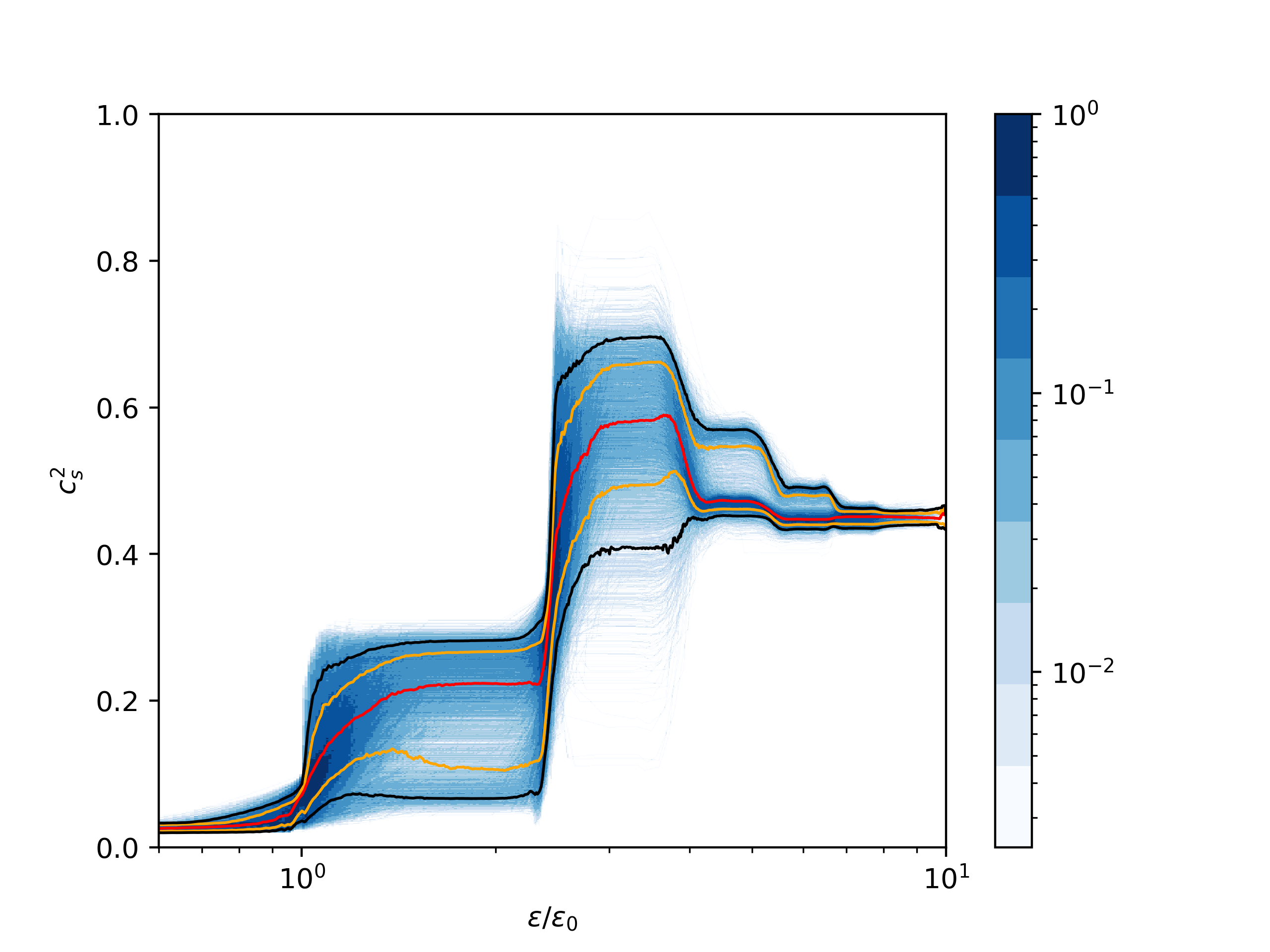}
	\caption{\small The PDF plot for the variation of $c_{s}^{2}$ against normalised energy density. The red coloured line shows the median, the orange coloured line depicts the $65\%$ confidence interval, and the black coloured lines denote the $90 \%$ confidence interval of our distribution.}
	\label{csPlot}
\end{figure}

We randomly selected $5,000$ EoSs from the remaining $20,000$, which were not used to train the NN. The TOV equations for these EoSs are solved. For each of these EoSs, $15$ MR points were chosen, which serve as input into our NN. The output gives us the six $c_{s}^{2}$ values corresponding to six segments of the EoSs. The obtained $c_{s}^{2}$ are then interpolated to construct an EoS termed `NN informed EoS'. This is repeated for each of the 5,000 EoSs, and we obtained 5,000 `NN informed EoSs'. We proceed with our analysis using our `NN informed EoSs'. In fig \ref{csPlot}, we show the probability density function (PDF) plot for the variation for the $c_{s}^{2}$ against $\epsilon / \epsilon_{0}$ for the NN informed EoS. For calculating the PDF \citep{Altiparmak_2022} we divide the entire plot into grids of resolution $1000 \times 1000$. Each cell is assigned a number equivalent to the number of curves passing through them, and then the entire grid is normalized by dividing them by the maximum value. The red line shows the median of the distribution. Orange and black lines depict the $65 \%$ and $90 \%$ confidence interval, respectively. We have trained our system till $10 \epsilon_0$ and obtained the speed of sound as an output from the NN. This is because the measurement constraints are only up to NS central densities, which for all the cases do not cross $10 \epsilon_0$. At low density for the first slice, $\epsilon_{0} \leq \epsilon \leq 2.5 \epsilon_{0}$, the speed of sound is around the conformal sound speed limit. This is expected as the speed of sound is directly related to the stiffness of the EoS. High sound speed at such densities would indicate that the EoS is substantially stiff in the outer regions for most stars, which is counter-intuitive. The nature of the speed of sound at intermediate densities shows non-monotonic behaviour, which other works also concluded \citep{Ma_2019, Ma_2021}.

The speed of sound sharply rises mainly at the end of the first slice (or, for some EoS, the beginning of the second slice), which is the central density of most intermediate stars. The peak is at around $3 \epsilon_0$ (for the median and $65\%$ and $90\%$ confidence level). The speed of sound then decreases and encounters a plateau beyond $5 \epsilon_0$, which is the central density of very massive stars (beyond $2.2$ solar mass). Therefore, the EoS slope rises very fast beyond $2$ times the saturation energy density (becomes stiff) and has a maximum around $3-4$ times the saturation energy density. The slope (or the stiffness) then decreases and becomes almost constant. 

\Scc{To connect this feature with the matter properties, the matter beyond the CET band becomes very stiff, and the speed of sound almost reaches the conformal value. The matter properties remain constant up to some density; however, it again stiffens beyond twice the saturation density and attains a peak around three times the saturation density, violating the conformal limit. This feature is necessary to explain the mass of recently detected massive pulsars \cite{Romani, Romani_2021, Demorest_2010, Altiparmak_2022,Reed2020, Traversi2022}. The peak speed of sound usually hints towards a phase change; therefore, as extra degrees of freedom appear, the sound speed decreases beyond the maximum value.} Therefore,
stars with central energy density $3-4$ times saturation density have a stiff central core and softer matter in the outer region. Extraordinarily massive stars (whose central energy density is beyond $4$ times saturation energy density) have a relatively softer core, intermediate stiff region and, again, a softer outer surface.

\begin{figure}
	\centering
	\begin{minipage}[ht]{0.5\textwidth} 
		\includegraphics[scale=0.60]{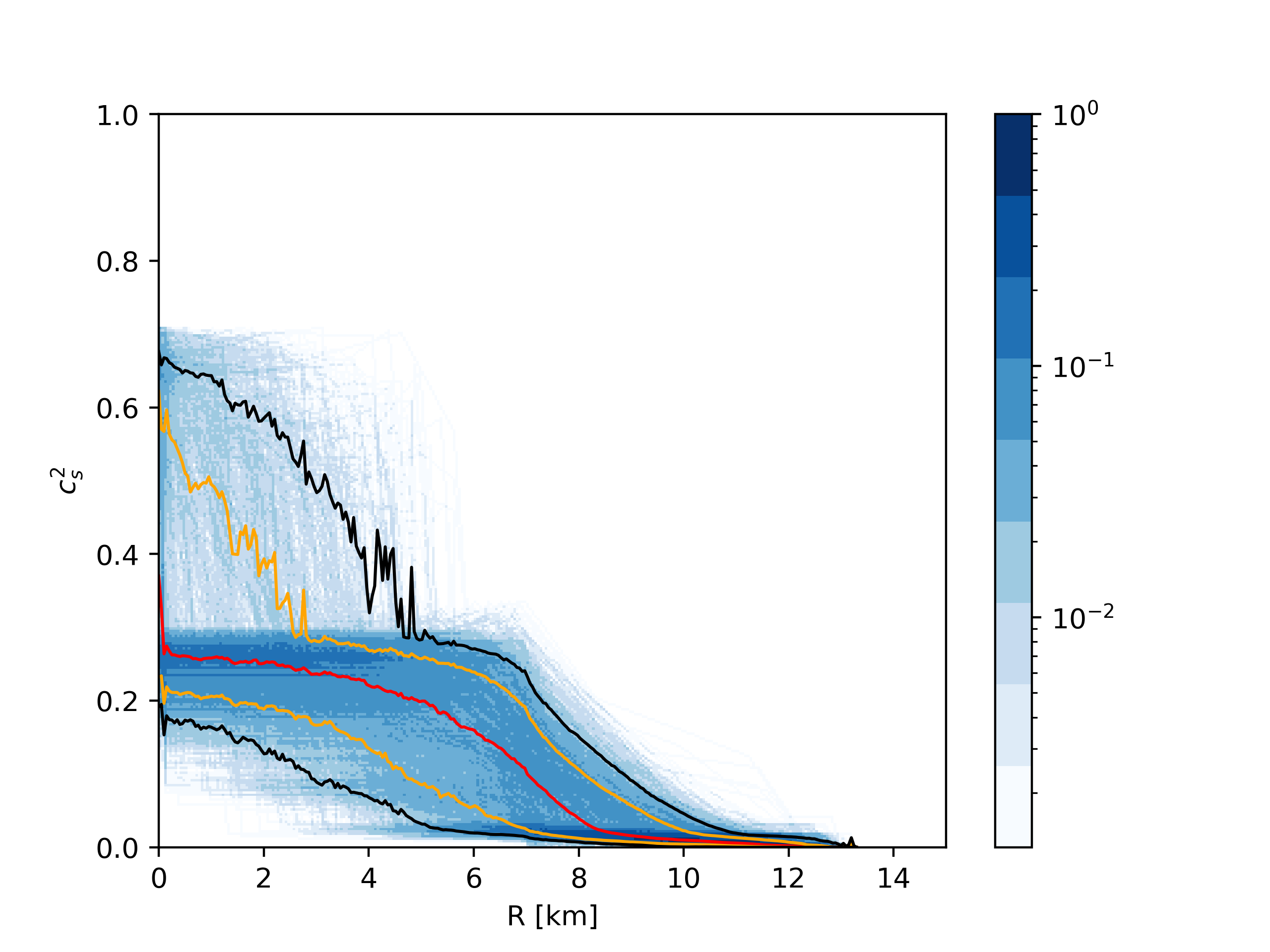}
	\end{minipage} 
	\begin{minipage}[ht]{0.5\textwidth} 
		\includegraphics[scale=0.60]{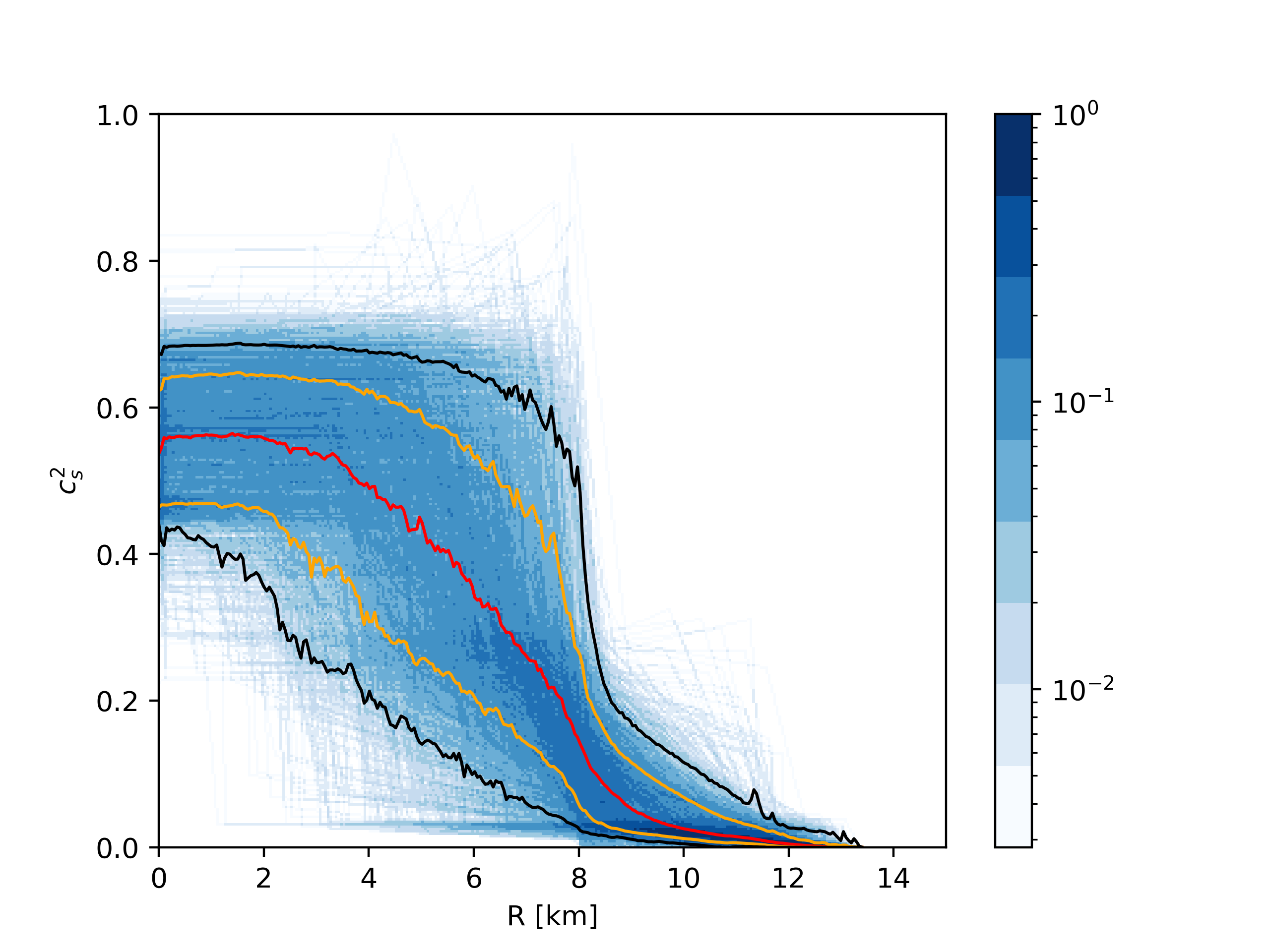}
	\end{minipage} 
	\caption{{\bf (Top):} \small The PDF plot for the variation of $c_s^{2}$ against the radius from the centre to the surface of a $1.6 M_{\odot}$ star is shown here. The median of the distribution is shown in red, with the 65\% confidence interval in orange, and the 90 \% is shown in black.{\bf (Bottom):} The PDF plot for the square of the speed of sound against the radius from the center to the surface of a $2.2 M_{\odot}$ star. The nomenclature remains the same.}
	\label{1_6}
\end{figure}

It would be insightful to plot the variation of $c_s^2$ inside the star to know more about the exact nature of the matter inside the star. 
In fig \ref{1_6} (top and bottom), we show the PDF plot for variation of the speed of sound inside the NS for stars of masses $1.6 M_{\odot}$ and $2.2 M_{\odot}$ respectively. 
For stars with $1.6$ solar mass, the central energy density is between $1.7-3$ times saturation density. \Scc{Most of the EoS for such stars have a speed of sound near the conformal value, and therefore, the median saturates near the conformal value at the core of the stars. However, for some EoS (whose stars have a core density around $3$ times saturation energy density), the speed of sound continues to rise. Therefore, the upper envelope of both $65\%$ and $90\%$ PDF shows an increasing nature. Moving towards the surface, we encounter a plateau region inside the stars, corresponding to the typical plateau region of fig \ref{csPlot} at the low-density part. Ultimately, near the surface, the sound speed falls to almost zero.} 

The central sound speed for $2.2 M_{\odot}$ star is different (fig \ref{1_6} bottom). As most of the stars with such mass have a central energy density beyond $3 \epsilon_0$ (the central energy density of such stars lies in the range $ 2.4 \epsilon_0 - 6 \epsilon_0$), the sound speed reaches its peak much before the central core. Therefore, the flattening (almost constant) of the sound speed is seen in the median, the $65\%$ and $90\%$ confidence region, even at relatively larger radial distances. As we go to the outer region, the plateau in the sound speed (slice one in fig \ref{csPlot}) is barely visible (which was very prominent for $1.6 M_{\odot}$ star).

\Scc{The distinct differences in the variation of $c_s^{2}$ for two masses, $1.6 M_{\odot}$ and $2.2 M_{\odot}$, are attributed to the fact that the central densities for them are different. The $1.6 M_{\odot}$ stars have a soft outer surface, and as we travel to the core, the matter stiffens. For most stars, the core stiffness till $4-6$ km is constant; however, for few stars, the matter continues to stiffen till the central core. On the other hand, $2.2 M_{\odot}$ star have a soft outer core and a stiff inner core till a radius of about $4-6$ km. The core stiffness of $1.6 M_{\odot}$ and $2.2 M_{\odot}$ differs considerably.}

\subsection{Trace anomaly} \label{sec:TA}

QCD in the massless limit shows conformal symmetry, and the expectation value of the trace of the energy-momentum tensor $<T_\mu^\mu>$ in the classical level becomes zero. The matter at finite temperature and/or chemical potential $<T_\mu^\mu>$ can be decomposed into a vacuum and matter part. For matter, the trace anomaly becomes $<T_\mu^\mu>=\epsilon-3p$. At finite temperature and low density, there is a sharp discontinuity in the normalized trace anomaly at the deconfinement transition temperature. This is due to the appearance of the gluonic degrees of freedom in the system. It would be interesting to check how the trace anomaly behaves for finite density and at low temperatures (in this case, $0$). As the trace anomaly depends on the thermodynamic properties of matter (namely pressure and energy density), it has a close connection with the speed of sound. 

It has been previously shown that the trace anomaly for an ideal fluid is given by \citep{Ecker_trace}.
\begin{equation}
<T_\mu^\mu> = \epsilon-3p.
\end{equation}
Sometimes, a measure of trace anomaly is defined as 
\begin{equation}
\Delta = \frac{<T_\mu^\mu>}{3\epsilon}=\frac{1}{3}-\frac{p}{\epsilon}.
\end{equation} 
\begin{figure}
	\centering
	\includegraphics[scale=0.60]{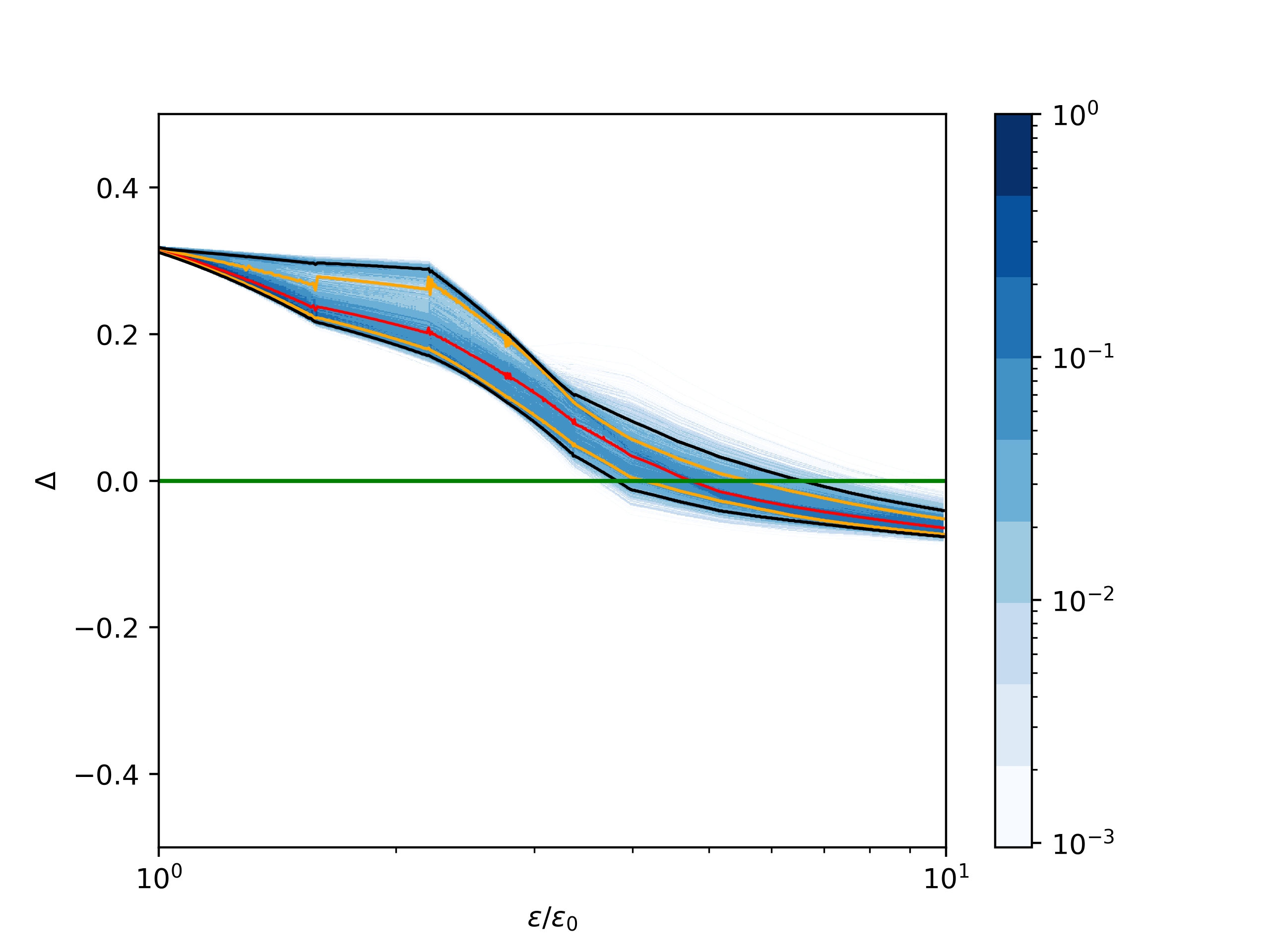}
	\caption{\small The PDF plot for the variation of $\Delta$ against the energy density is shown here. The median of the distribution is shown in red, with the 65\% confidence interval in orange, and the 90 \% is shown in black. }
	\label{trace}
\end{figure} 

Fig \ref{trace} shows the PDF plot of the measure of trace anomaly with energy density for the NN informed EoS. 
The median starts from a high value (near about $1/3$) and then decreases as we move to high density. The $\Delta$ reaches zero (the conformal value) between $4-7$ times the saturation density and continues to decrease, taking negative values. The minimum value of the trace anomaly in the region of interest (till $10 \epsilon_0$) is $\approx -0.076$ (for a $90 \%$ confidence interval); however, this is not the troughs. Although we encounter a peak in the speed of sound, we do not find any troughs for trace anomaly in the region where the central density of neutron stars lies. The trace anomaly has a minimum around $18 \epsilon_0$ (median) and then increases (fig \ref{fig:trace-NoPQCD} in the \ref{app:pQCD}) when plotted till the asymptotically high-density range. Ultimately, it tends to the conformal value $0$ at very high densities. It approaches zero from the positive side. 

\begin{figure}
	\centering
	\begin{minipage}[ht]{0.5\textwidth} 
		\includegraphics[scale=0.60]{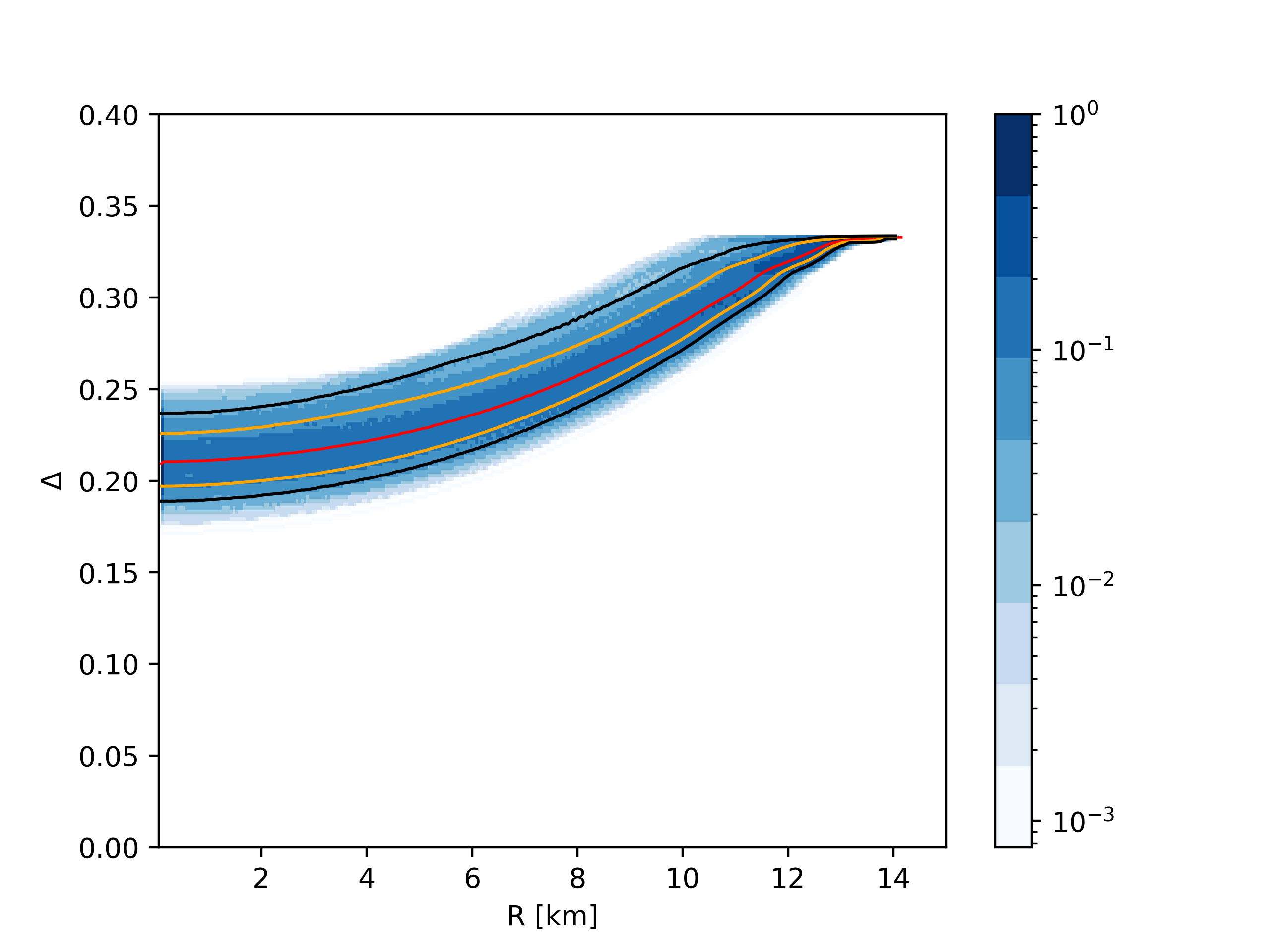}
	\end{minipage} 
	\begin{minipage}[ht]{0.5\textwidth} 
		\includegraphics[scale=0.60]{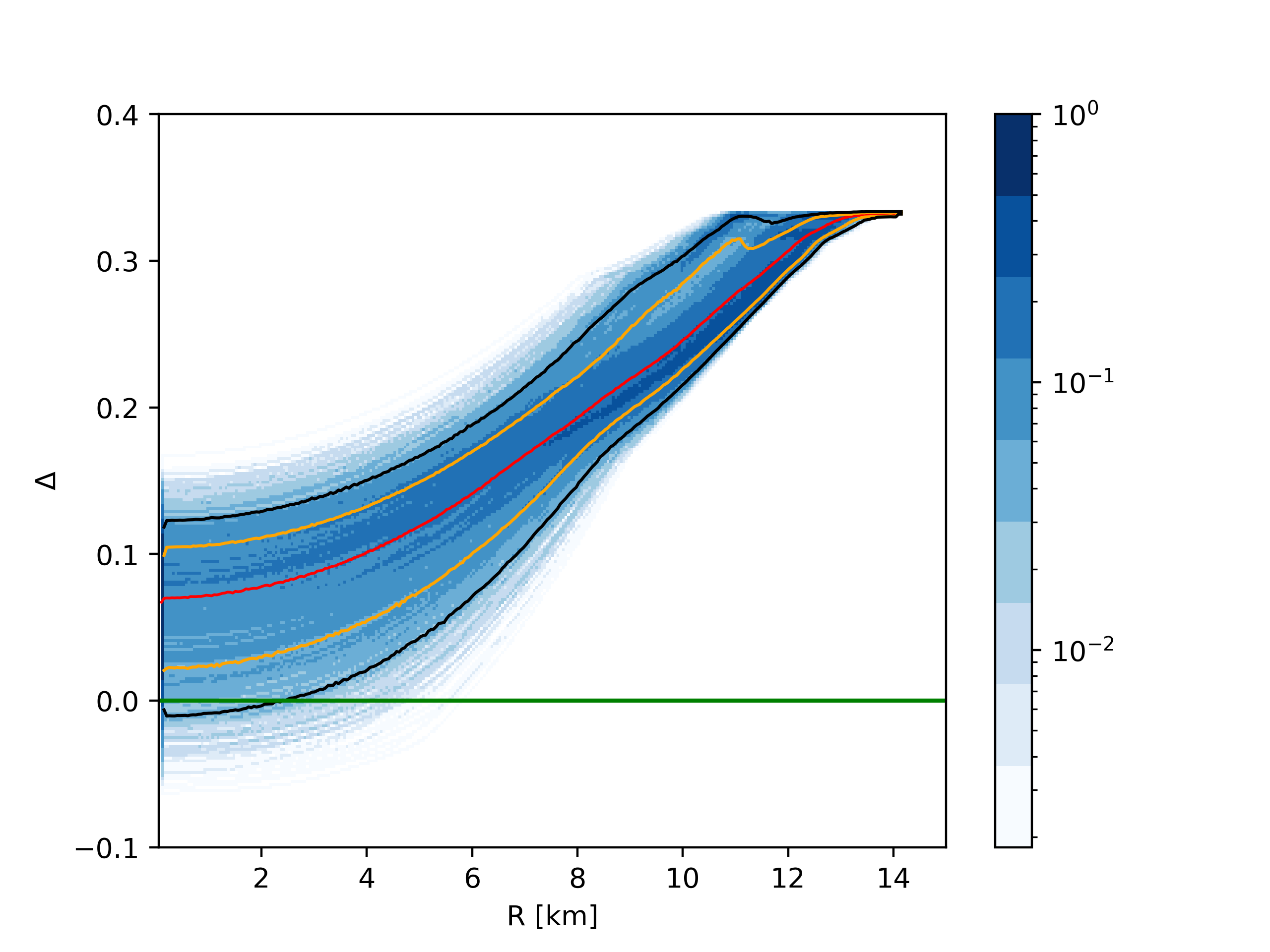}
	\end{minipage} 
	\caption{{\bf (Top):} \small The PDF plot for the variation of $\Delta$ against the radius from the centre to the surface of a $1.6 M_{\odot}$ star is shown here. The median of the distribution is shown in red, with the 65\% confidence interval shown in orange, and the 90 \% is shown in black.{\bf (Bottom):} The PDF plot for the variation of $\Delta$ against the radius from the centre to the surface of a $2.2 M_{\odot}$ star. The nomenclature remains the same.}
	\label{1_6-trace}
\end{figure}

\Scc{In fig \ref{1_6-trace}, we show how the measure of trace anomaly varies for $1.6 M_{\odot}$ and $2.2M_{\odot}$ star from centre to the surface. The results for $1.6 M_{\odot}$ stars show that $\Delta > 0$ throughout the star from centre to the surface \Comm{with a minimum value of $0.185$ at $90\%$ confidence interval}, unlike that of the $2.2 M_{\odot}$ stars \Comm{which shows a negative minimum value of $-0.012$ at the same confidence interval}. For the higher-mass stars, we see that the median of the distribution does not cross zero. However, a finite number of EoSs show $\Delta < 0$ behaviour inside the star. 
The trace anomaly approaches zero and becomes slightly negative for maximal densities of neutron stars. At very low density, the pressure is negligible compared to the energy density; therefore, near the star's surface, the ratio $p/ \epsilon \rightarrow 0$. This makes $\Delta \rightarrow 1/3$. However, the ratio $p/ \epsilon$ becomes maximum close to the star's centre. This accounts for low values of $\Delta$ at the core of the stars. For massive stars, since the value of $p/ \epsilon$ is higher, hence we observe a lower value of $\Delta$ compared to low-mass stars. The conformal limit predicts the trace anomaly to go to zero. However, this is only true at asymptotically high densities where conformal symmetry is restored, and we have weakly interacting quark matter. A negative value of trace anomaly may imply that at the core of a neutron star, we may not have purely weakly interacting conformal quark matter and instead have near-conformal matter  \cite{Annala_2023, Fujimoto_Trace, Cai}}.

\section{Summary and Conclusion} \label{sec:Summary}

In this paper, we have studied how the speed of sound and trace anomaly varies with density inside a neutron star. As the intermediate density range is beyond the scope of experiments and direct theoretical calculation, we probe the density range with NS observations and constraints. We construct a family of agnostic EoS by maintaining thermodynamic stability and the speed of sound bound. Randomly selecting $20000$ EoS, we construct the mass-radius curve for these EoS. 
Using Gaussian marginalisation of neutron star mass and radius from several pulsar observations, we introduce error in the data and obtain $2 \times10^6$ data set; each set having $15$ pairs of data points. This serves as a training set for the NN, and we train our NN, minimising the loss function. 
A new set of $5000$ EoS is then used to obtain $5000$ mass-radius sequences. Again, we randomly select $15$ points from each of the M-R sequences, which serve as input for the trained neural network to obtain the speed of sound as an output. 
This speed of sound is interpolated to reconstruct the equation of state. 

\Scc{The study shows that beyond the CET band, sound speed stiffens sharply, attains a value close to the conformal value, and remains flat until around twice the nuclear saturation density. Beyond twice the saturation energy density, it again stiffens sharply and attains a peak breaking the conformal bound. This feature is necessary to address the mass of massive pulsars. Beyond the maximum value, it falls off, and we encounter a peak that hints towards a new phase and generation of extra degrees of freedom.}

\Scc{The speed of sound analysis inside two stars of masses $1.6 M_{\odot}$ and $2.2 M_{\odot}$ points to the fact that most of the intermediate-mass stars have a stiffer inner core and softer outer surface. The stiffness of the inner core is close to the conformal value. However, the stiffness from the surface to the core monotonically increases for a few intermediate-mass stars. For very massive stars, the star can be divided into a stiff inner core, stiffness beyond the conformal limit, and a softer outer layer. Massive stars' violation of conformal value hints towards the appearance of a new phase at their core.}

The regime between saturation energy density to $8-10$ times the saturation energy density is constrained by the astrophysical observation of neutron stars.
The nature of EoS in this regime reveals a very robust fact that the speed of sound has to rise in the intermediate density range, attain a peak, and fall off before $8-10$ times saturation density. This is consistent with other works \citep{Ecker_2022, Fujimoto_Trace}. However, the position of the peak is still ambiguous. With our NN approach, even for very massive stars $2.2 M_{\odot}$, the speed of sound does not decrease significantly at the centre, which is in contrast from some previous calculations \citep{Ecker_2022}. Analysing more massive stars is also not viable as it may not be astrophysically stable.\Scc{The change in speed of sound is usually associated with a change in matter properties of the stars \cite{Bedaque2015,Tews2018,Ferrer2023}. Hadronic matter typically exhibits $c_s^{2} \le 1/3$ at lower densities, but at higher densities, they can cross this limit, which most likely indicates a change in matter properties \cite{Reed2020}. Thus, the presence of a maximum of $c_s^{2}$ for higher mass stars hints towards the possibility of higher mass stars being hybrid stars (possibly having a quark core surrounded by hadronic matter). The absence of this maxima at lower mass stars hints towards the idea that lower mass stars are indeed composed of hadronic matter.}

The NN informed EoSs data for the conformal anomaly also shows non-monotonicity in their behaviour. It goes to zero at the asymptotic-density limit; however, the $\Delta \ge 0$ condition is not maintained in the intermediate density region. It goes to negative values even inside the NS. \Scc{This behaviour was also observed at the interior of the stars where we saw that a distinct change in sign is observed for a star of $2.2 M_{\odot}$. The conformal limit predicts that $c_s^2=1/3$ and the trace anomaly goes to zero. However, this is only true at asymptotically high densities where conformal symmetry is restored, and we have weakly interacting quark matter. Therefore, a negative value of trace anomaly may imply that at the cores of neutron stars, we may not have purely weakly interacting conformal quark matter. Therefore, at the cores, it can have other form of near-conformal matter (not necessarily quark matter). Therefore, trace anomaly being negative suggests that cores of massive neutron stars may contain near-conformal matter and not exactly conformal quark matter}

We have used an optimal segmentation of the density range; having higher segmentation does not affect our results significantly. One can also do similar calculations directly relating the EoS with the mass and radius of the star with a more sophisticated model like an auto-encoder. Stricter mass-radius measurements from the NICER experiments involving more pulsars would make such studies more robust. 

\section*{Acknowledgements}

The authors would like to thank IISER Bhopal for providing the infrastructure to carry out the research. SC would like to acknowledge the Prime Minister's Research Fellowship (PMRF), Ministry of Education Govt. of India, for a graduate fellowship. The authors are grateful to the Science and Engineering Research Board (SERB), Govt. of India for monetary support in the form of Core Research Grant (CRG/2022/000663). The authors would like to thank Sayan Roy, Kamal Krishna Nath, Shamim Haque and Pratik Thakur for their helpful discussions.

\section*{Data Availability}
The data can be availed on reasonable request to the corresponding author.

\appendix

\section{Impact of pQCD constraints}\label{app:pQCD}
\begin{figure}
    \centering
\begin{minipage}[ht]{0.5\textwidth}        \includegraphics[scale=0.6]{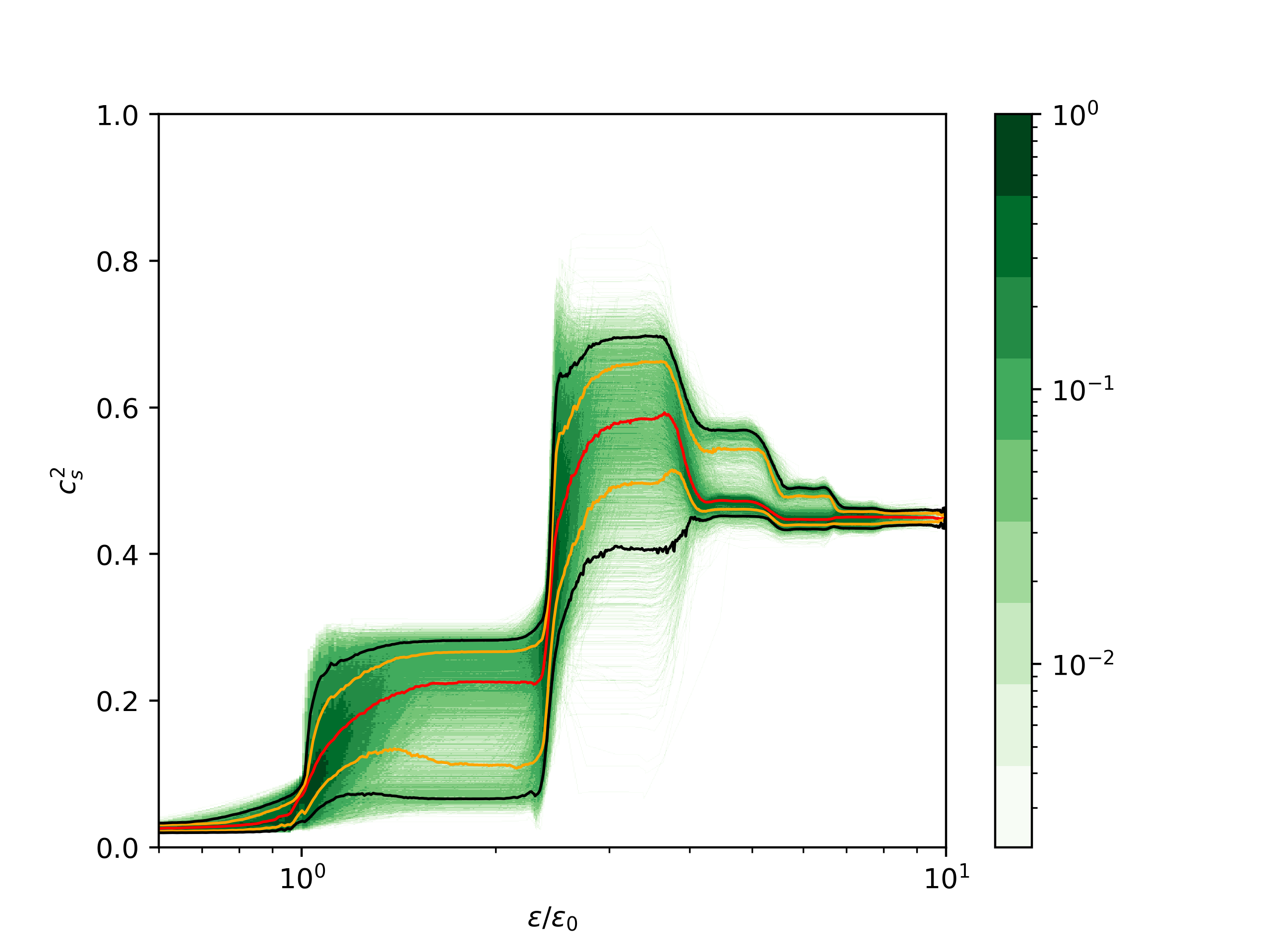}

    \end{minipage}

\begin{minipage}[ht]{0.5\textwidth}    \includegraphics[scale=0.6]{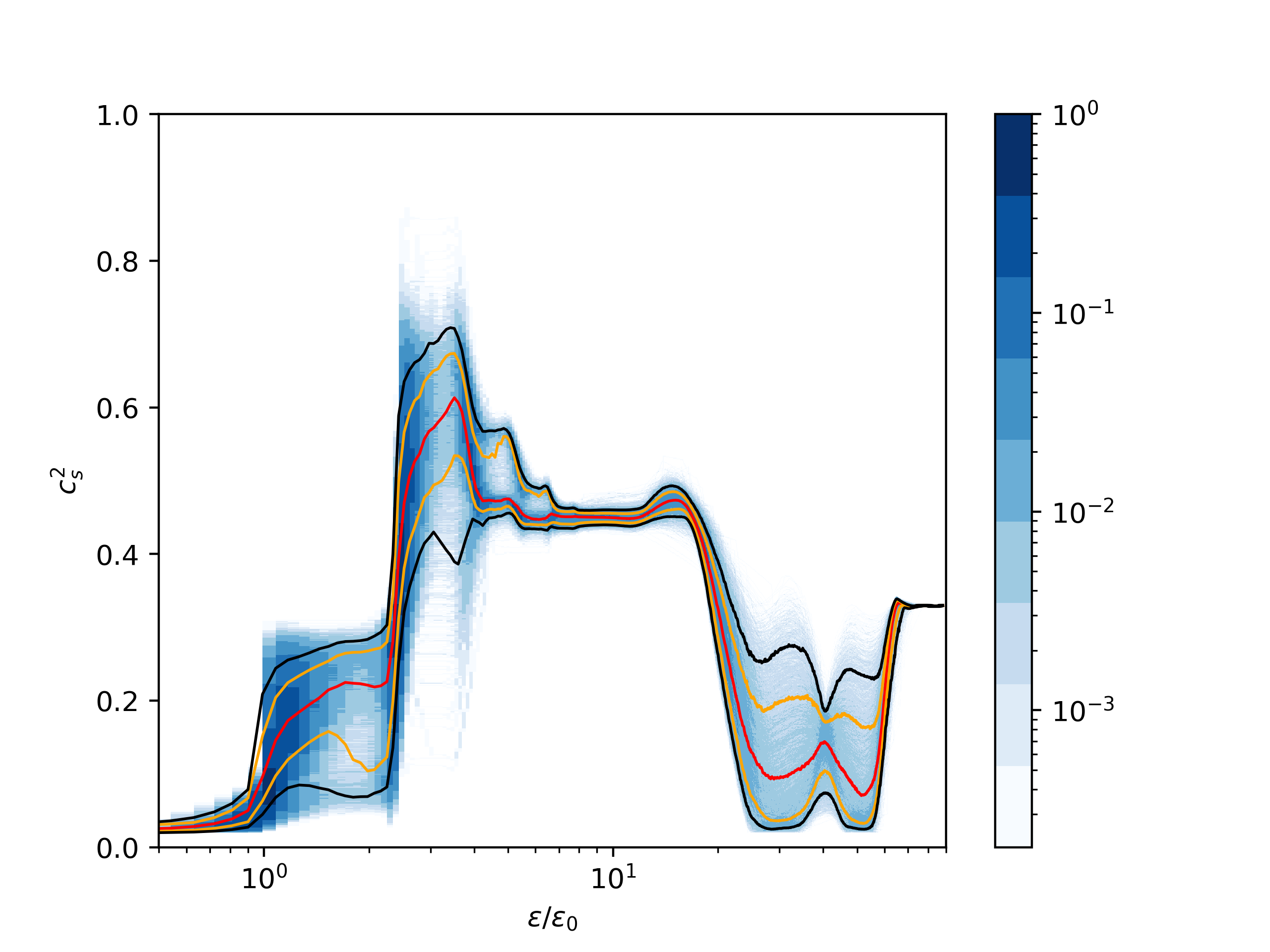}
\end{minipage}
    
    \caption{\textbf{(Top)}: The PDF distribution for the variation of $c_s^{2}$ with normalized energy density for EoSs shown in green (without pQCD constraints). The red coloured line depicts the median, the orange coloured line depicts the region of 65 $\%$ confidence, and the black region depicts the 90 $\%$ confidence region. \textbf{(Bottom):} The PDF for the EoSs with pQCD constraints are shown in blue. Unlike the other plot, this plot shows the variation till the pQCD limit (where the pQCD constraint has been imposed). The other nomenclatures remain the same as previous. }
    \label{fig:cs2-NoPQCD}
\end{figure}

\Scc{In this section, we discuss the possible implications of the pQCD constraints on our analysis. Earlier, we had discussed that since the central density of our stars does not cross $10 \epsilon_0$ hence, beyond this density, we cannot comment on the nature of EoS. To further strengthen our argument, we perform the analysis equivalent to fig \ref{csPlot} and fig \ref{trace}, the only difference being that for one set, we restrict our analysis till $10 \epsilon_0$ and for the other till $90 \epsilon_0$. This would show whether the pQCD limit has any influence in determining the variation of $c_s^{2}$ and $\Delta$ for our case. For both these scenarios, we plot them till $10 \epsilon_0$ for comparison. From fig \ref{fig:cs2-NoPQCD}, we see that the EoSs without pQCD constraint (shown in green) and the EoSs with pQCD constraint (shown in blue) appear similar with no prominent characteristic indicating any influence of the constraints on the EoSs.}

\begin{figure}
    \centering
    \includegraphics[scale=0.5]{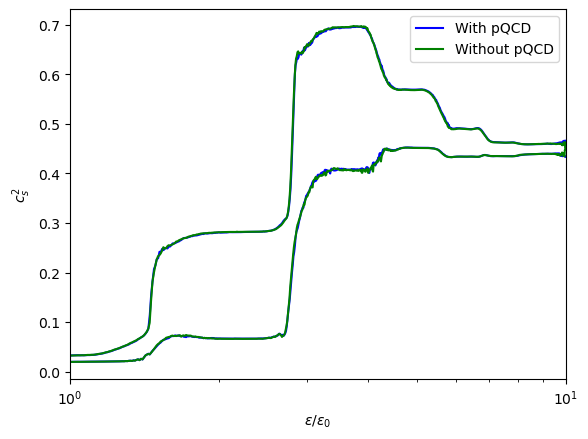}
    \caption{Figure depicts theregion corresponding to the $65\%$ contours in fig \ref{fig:cs2-NoPQCD} (orange lines) till $10 \epsilon_0$.}
    \label{fig:cs2-Compare}
\end{figure}

\Scc{To have a much clearer picture, we show the shaded region corresponding to the $65 \%$ confidence region corresponding to fig \ref{fig:cs2-NoPQCD} in fig \ref{fig:cs2-Compare}. From fig \ref{fig:cs2-Compare}, we see that there is no visible impact of the pQCD constraints on the $c_s^2$, which was expected as we only train our NN till energy density of $10 \epsilon_0$.}

\begin{figure}
    \centering
\begin{minipage}[ht]{0.5\textwidth}        \includegraphics[scale=0.55]{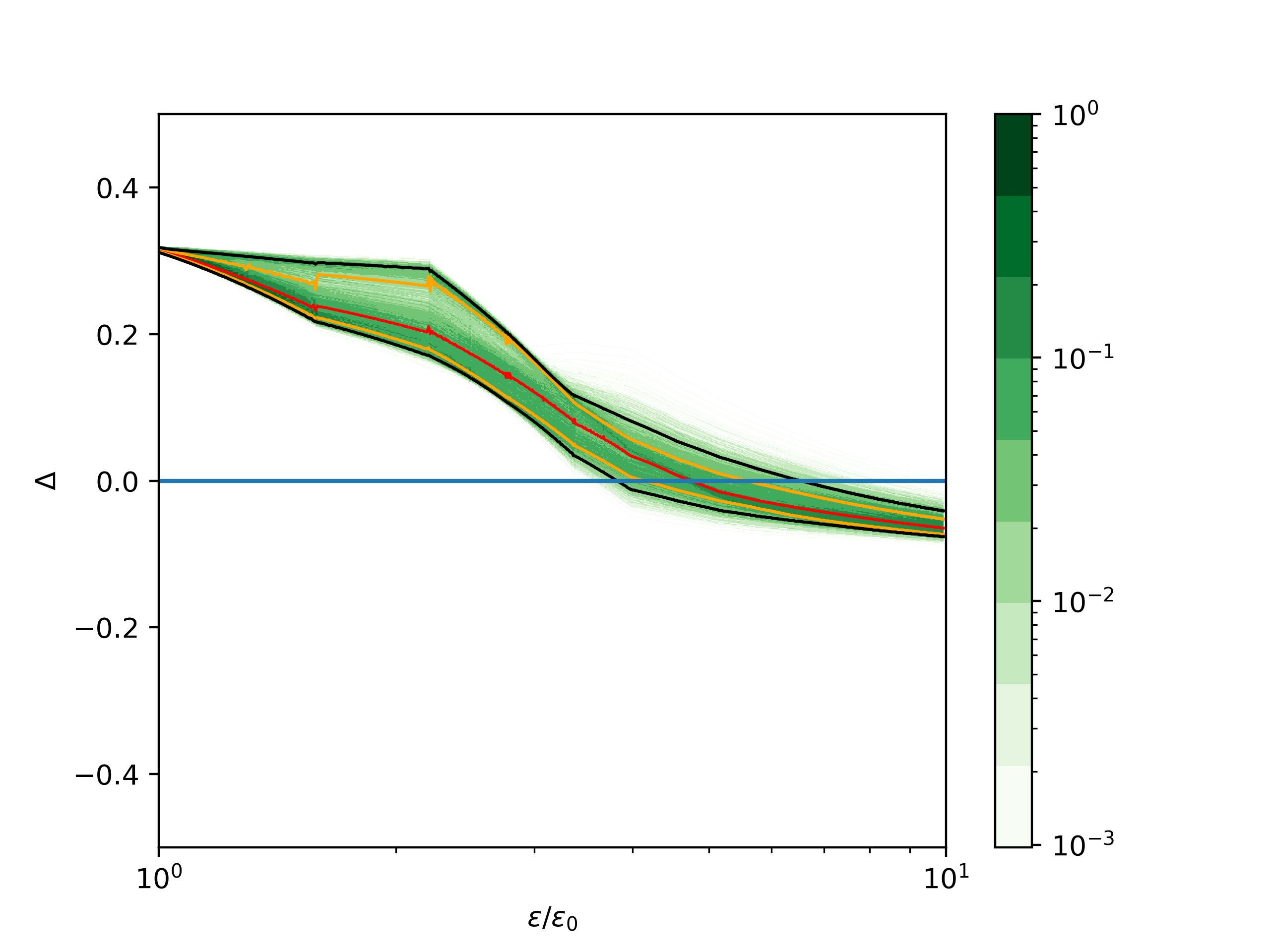}

    \end{minipage}

\begin{minipage}[ht]{0.5\textwidth}    \includegraphics[scale=0.55]{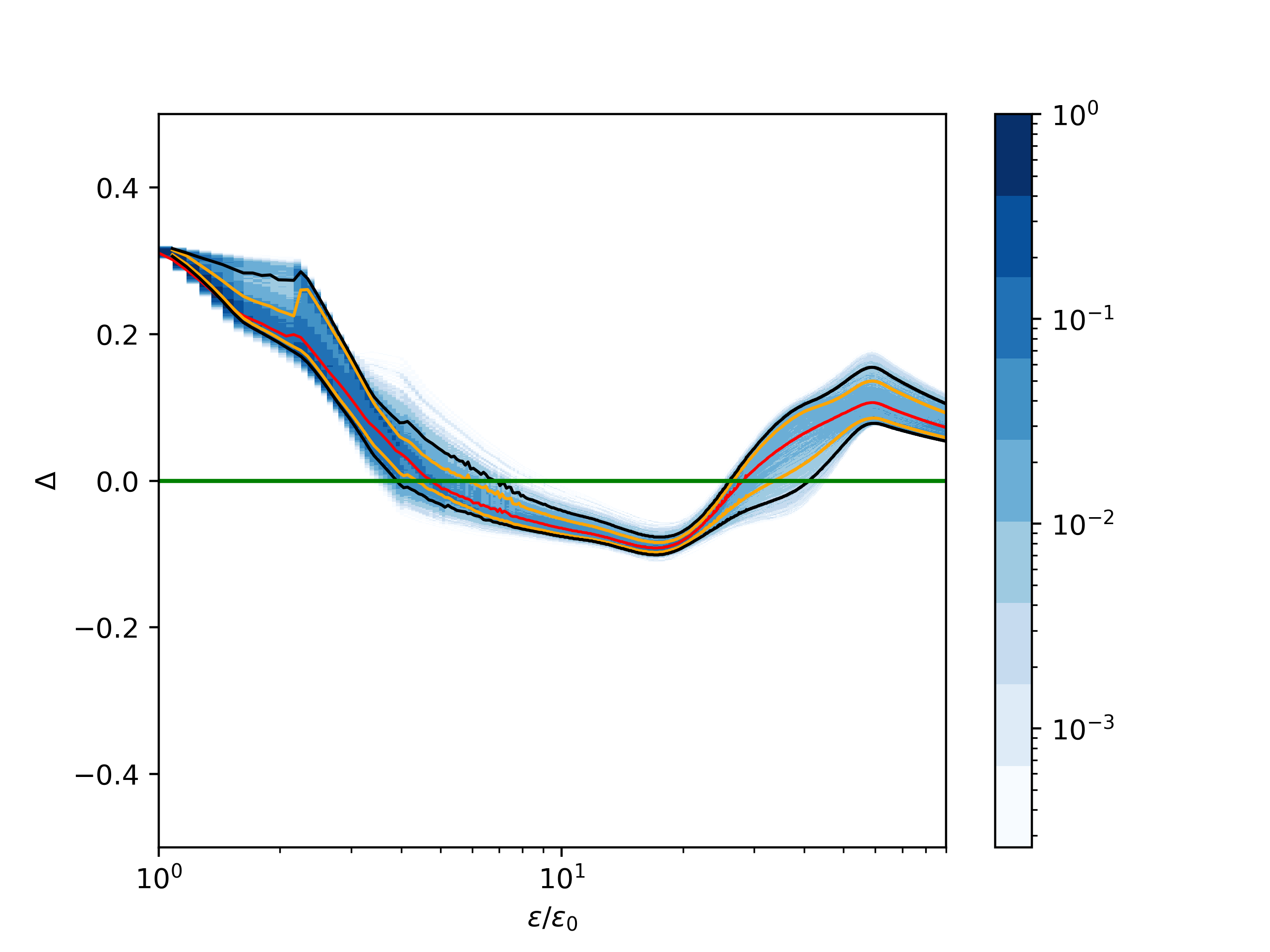}
\end{minipage}
    
    \caption{\textbf{(Top)}: The PDF distribution for the variation of the measure of trace anomaly $\Delta$ with normalized energy density for EoSs shown in green (without pQCD constraints). The red coloured line depicts the median, the orange coloured line depicts the region of 65 $\%$ confidence, and the black region depicts the 90 $\%$ confidence region. \textbf{(Bottom):} The PDF for the EoSs with pQCD constraints are shown in blue till $90 \epsilon_0$. The other nomenclatures remain the same as previous.}
    \label{fig:trace-NoPQCD}
\end{figure}

\begin{figure}
    \centering
    \includegraphics[scale=0.5]{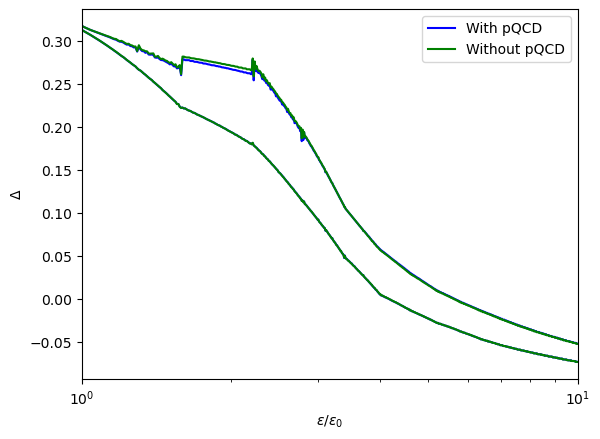}
    \caption{Figure depicts the region corresponding to the $65\%$ contours in fig \ref{fig:trace-NoPQCD} (orange lines) till $10 \epsilon_0$.}
    \label{fig:trace-Compare}
\end{figure}

\Scc{We perform the same analysis for the measure of trace anomaly. This is shown in fig \ref{fig:trace-NoPQCD} and fig \ref{fig:trace-Compare}. The results again draw a similar conclusion that the pQCD constraints do not affect our analysis, as NSs are not formed at extremely high densities.}

\section{Comparisons of different optimizers and activation functions} \label{appx:opt}

\Scc{In this section, we show the effects of different optimizers and activation functions on our analysis. We use 10,000 EoSs and the optimizers - Adam, Nestrov-Adam, RMSprop, and SGD. We also change the activation functions between ReLU, GELU, SELU, Mish, softmax, and swish for each one. Table \ref{table_opt} presents the training loss after 1000 epochs for the same 10,000 EoSs. The minimum loss is highlighted in bold in the table.}

\begin{table*}
     \centering
      \caption{ Table showing the minimum training loss after 1000 epochs for same sets of 10,000 EoSs. The minimum value is marked in bold.}
     \label{table_opt}
     \begin{tabular}{@{\hspace{0.8cm}}c @{\hspace{0.8cm}}c @{\hspace{0.8cm}}c @{\hspace{0.8cm}}c @{\hspace{0.8cm}}c @{\hspace{0.8cm}}c @{\hspace{0.8cm}}c}
     \hline
     \hline
          & ReLU & GELU & SELU & Mish & Softmax & Swish \\ \hline  \\
         Adam & 0.027976 & 0.027839 & 0.028231 & 0.027882 & 0.029248 & 0.027707 \\  \\ \hline \\
         Nestrov-Adam &  0.028356 & 0.028068 & 0.028254 & \textbf{0.027551}& 0.087289 &0.027776   \\ \\  \hline \\
          RMSprop & 0.028107 & 0.027911 & 0.027972 & 0.027926 & 0.028012& 0.027943 \\ \\  \hline \\
          SGD & 0.028129 & 0.028151 & 0.078673 &0.027860& 0.102064 & 0.028262\\  \\
           \\ \hline\hline \\
     \end{tabular}
    
 \end{table*}


\clearpage
\newpage

\begin{thebibliography}{12}



\bibitem{Valentim}
Valentim, R., Rangel, E., Horvath, J. E.
\newblock {On the mass distribution of neutron stars}.
\newblock {\em Monthly Notices of the Royal Astronomical Society}, \textbf{414} (2), 1427-1431, 2011.
\newblock \url{https://doi.org/10.1111/j.1365-2966.2011.18477.x}

\bibitem{Witten}
E. Witten,
\textit{Phys. Rev. D} \textbf{30}, 272--285 (1984).

\bibitem{Shuryak}
E. V. Shuryak,
\textit{Phys. Rep.} \textbf{61}, 71--158 (1980).


\bibitem{Riley_2019}
Riley, T. E., Watts, A. L., Bogdanov, S., et al.
\newblock {A NICER View of PSR J0030+0451: Millisecond Pulsar Parameter Estimation}.
\newblock {\em ApJL}, \textbf{887} (1), L21, 2019.
\newblock \url{https://doi.org/10.3847/2041-8213/ab481c}

\bibitem{Riley_2021}
Riley, T. E., Watts, A. L., Ray, P. S., et al.
\newblock {A NICER View of the Massive Pulsar PSR J0740+6620 Informed by Radio Timing and XMM-Newton Spectroscopy}.
\newblock {\em The Astrophysical Journal Letters}, \textbf{918} (2), L27, 2021.
\newblock \url{https://dx.doi.org/10.3847/2041-8213/ac0a81}


\bibitem{Miller_2019}
Miller, M. C., Lamb, F. K., Dittmann, A. J., et al.
\newblock {PSR J0030+0451 Mass and Radius from NICER Data and Implications for the Properties of Neutron Star Matter}.
\newblock {\em The Astrophysical Journal Letters}, \textbf{887} (1), L24, 2019.
\newblock \url{https://dx.doi.org/10.3847/2041-8213/ab50c5}

\bibitem{Miller}
M. C. Miller et al., "The Radius of PSR J0740+6620 from NICER and XMM-Newton Data," Astrophys. J. Lett. \textbf{918}(2), L28 (2021), doi:10.3847/2041-8213/ac089b.


\bibitem{Abbot}
Abbott, B. P., Abbott, R., Abbott, T. D., et al. (LIGO Scientific Collaboration and Virgo Collaboration).
\newblock {GW170817: Observation of Gravitational Waves from a Binary Neutron Star Inspiral}.
\newblock {\em Physical Review Letters}, 119(16):161101, 2017.


\bibitem{Hebeler_2013}
K. Hebeler et al., "Equation of State and Neutron Star Properties Constrained by Nuclear Physics and Observation," Astrophys. J. \textbf{773} (1), 11 (2013), doi:10.1088/0004-637X/773/1/11.

\bibitem{Lattimer}
J. M. Lattimer \& M. Prakash,
\textit{Science} \textbf{304}, 536-542 (2004).


\bibitem{deForcrand}
P. de Forcrand,
\textit{PoS LAT2009}, \textbf{010} (2010).


\bibitem{Most}
E. R. Most, L. R. Weih, L. Rezzolla, \& J. Schaffner-Bielich,
\textit{Phys. Rev. Lett.} \textbf{120}, 261103 (2018).


\bibitem{Annala}
E. Annala, T. Gorda, A. Kurkela, \& A. Vuorinen,
\textit{Phys. Rev. Lett.} \textbf{120}, 172703 (2018).


\bibitem{Zhang_2018}
N.-B. Zhang, B.-A. Li, \& J. Xu,
\textit{The Astrophysical Journal} \textbf{859}, 90 (2018).


\bibitem{Bedaque}
Bedaque, P., Steiner, A. W.
\newblock {Sound Velocity Bound and Neutron Stars}.
\newblock {\em Phys. Rev. Lett.}, \textbf{114} (3), 031103, 2015.
\newblock \url{https://link.aps.org/doi/10.1103/PhysRevLett.114.031103}

\bibitem{Komoltsev}
Komoltsev, O., Kurkela, A.
\newblock {How Perturbative QCD Constrains the Equation of State at Neutron-Star Densities}.
\newblock {\em Phys. Rev. Lett.}, \textbf{128} (20), 202701, 2022.
\newblock \url{https://link.aps.org/doi/10.1103/PhysRevLett.128.202701}


\bibitem{Kojo2021}
Kojo, T.
\newblock {QCD equations of state and speed of sound in neutron stars}.
\newblock {\em AAPPS Bulletin}, \textbf{31} (1), 11, Apr 2021.
\newblock \url{https://doi.org/10.1007/s43673-021-00011-6}

\bibitem{Altiparmak_2022}
S. Altiparmak, C. Ecker, \& L. Rezzolla, "On the Sound Speed in Neutron Stars," Astrophys. J. Lett. \textbf{939}(2), L34 (2022), doi:10.3847/2041-8213/ac9b2a.




\bibitem{Fujimoto2021}
Y. Fujimoto, K. Fukushima, and K. Murase, "Extensive studies of the neutron star equation of state from the deep learning inference with the observational data augmentation," J. High Energ. Phys. 2021.3 (2021): \textbf{273}, doi:10.1007/JHEP03(2021)273.


\bibitem{Fujimoto2018}
Y. Fujimoto, K. Fukushima, and K. Murase, "Methodology study of machine learning for the neutron star equation of state," Phys. Rev. D 98.2 (2018): \textbf{023019}, doi:10.1103/PhysRevD.98.023019.


\bibitem{Fujimoto_2020}
Fujimoto, Y., Fukushima, K., Murase, K.
\newblock {Mapping neutron star data to the equation of state using the deep neural network}.
\newblock {\em Phys. Rev. D}, \textbf{101} (5), 054016, 2020.
\newblock \url{https://link.aps.org/doi/10.1103/PhysRevD.101.054016}

\bibitem{Morawski}
Morawski, F., Bejger, M.
\newblock {Neural network reconstruction of the dense matter equation of state derived from the parameters of neutron stars}.
\newblock {\em A\&A}, \textbf{642}, A78, 2020.
\newblock \url{https://doi.org/10.1051/0004-6361/202038130}

\bibitem{Ferreira_2021}
Ferreira, M., Providência, C.
\newblock {Unveiling the nuclear matter EoS from neutron star properties: a supervised machine learning approach}.
\newblock {\em Journal of Cosmology and Astroparticle Physics}, 2021(07), \textbf{011}.
\newblock \url{https://dx.doi.org/10.1088/1475-7516/2021/07/011}

\bibitem{Krastev}
Krastev, P. G.
\newblock {Translating Neutron Star Observations to Nuclear Symmetry Energy via Deep Neural Networks}.
\newblock {\em Galaxies}, 10(1), \textbf{16}, 2022.
\newblock \url{https://www.mdpi.com/2075-4434/10/1/16}


\bibitem{Traversi_2020}
Silvia Traversi and Prasanta Char,
\textit{The Astrophysical Journal} \textbf{905}, 9 (2020).
\newblock doi: \href{https://doi.org/10.3847/1538-4357/abbfb4}{10.3847/1538-4357/abbfb4}.

\bibitem{Soma_2022}
Soma, S., Wang, L., Shi, S., Stöcker, H., Zhou, K.
\newblock {Neural network reconstruction of the dense matter equation of state from neutron star observables}.
\newblock {\em Journal of Cosmology and Astroparticle Physics}, 2022(08), \textbf{071}.
\newblock \url{https://dx.doi.org/10.1088/1475-7516/2022/08/071}

\bibitem{Ferreira}
M\'arcio Ferreira, Val\'eria Carvalho, and Constan\ifmmode \mbox{\c{c}}\else \c{c}\fi{}a Provid\^encia,
\textit{Phys. Rev. D} \textbf{106}, \textbf{103023} (2022).
\newblock doi: \href{https://doi.org/10.1103/PhysRevD.106.103023}{10.1103/PhysRevD.106.103023}.




\bibitem{Raithel_2016}
Carolyn A. Raithel, Feryal Özel, and Dimitrios Psaltis,
\textit{The Astrophysical Journal} \textbf{831}, 44 (2016).
\newblock doi: \href{https://doi.org/10.3847/0004-637X/831/1/44}{10.3847/0004-637X/831/1/44}.


\bibitem{Raithel_2022}
Carolyn A. Raithel and Elias R. Most,
\textit{Phys. Rev. D} \textbf{108}, 023010 (2023).
\newblock doi: \href{https://doi.org/10.1103/PhysRevD.108.023010}{10.1103/PhysRevD.108.023010}.
\newblock eprint: \href{https://arxiv.org/abs/2208.04295}{arXiv:2208.04295} [astro-ph.HE].

\bibitem{Lindblom_2010}
Lee Lindblom,
\textit{Phys. Rev. D} \textbf{82}, 103011 (2010).
\newblock doi: \href{https://doi.org/10.1103/PhysRevD.82.103011}{10.1103/PhysRevD.82.103011}.

\bibitem{Baym1971}
Gordon Baym, Christopher Pethick, and Peter Sutherland, "The Ground State of Matter at High Densities: Equation of State and Stellar Models," Astrophys. J. \textbf{170} (1971): 299, doi:10.1086/151216.



\bibitem{Keller}
J. Keller et al., "Neutron matter at finite temperature based on chiral effective field theory interactions," Phys. Rev. C 103(5), \textbf{055806} (2021), doi:10.1103/PhysRevC.103.055806.


\bibitem{Epelbaum2009}
Epelbaum, E., Krebs, H., Lee, D., Mei{\ss}ner, U.-G.
\newblock {Ground-state energy of dilute neutron matter at next-to-leading order in lattice chiral effective field theory}.
\newblock {\em The European Physical Journal A}, \textbf{40} (2), 199-213, May 2009.
\newblock \url{https://doi.org/10.1140/epja/i2009-10755-0}

\bibitem{Tews}
Tews, I., Kr\"uger, T., Hebeler, K., Schwenk, A.
\newblock {Neutron Matter at Next-to-Next-to-Next-to-Leading Order in Chiral Effective Field Theory}.
\newblock {\em Phys. Rev. Lett.}, \textbf{110} (3), 032504, Jan 2013.
\newblock \url{https://link.aps.org/doi/10.1103/PhysRevLett.110.032504}



\bibitem{Gorda_2018}
T. Gorda et al., "Next-to-Next-to-Next-to-Leading Order Pressure of Cold Quark Matter: Leading Logarithm," Phys. Rev. Lett. \textbf{121}(20), 202701 (2018), doi:10.1103/PhysRevLett.121.202701.





\bibitem{Kurkela_2010}
A. Kurkela, P. Romatschke, \& A. Vuorinen, "Cold quark matter," Phys. Rev. D 81(10), \textbf{105021} (2010), doi:10.1103/PhysRevD.81.105021.


\bibitem{TOV}
Oppenheimer, J. R., Volkoff, G. M.
\newblock {On Massive Neutron Cores}.
\newblock {\em Phys. Rev.}, \textbf{55} (4), 374-381, Feb 1939.
\newblock \url{https://link.aps.org/doi/10.1103/PhysRev.55.374}

\bibitem{Romani}
R.~W.~Romani, D.~Kandel, A.~V.~Filippenko, T.~G.~Brink, and W.~K.~Zheng, 
"PSR J0952 0607: The Fastest and Heaviest Known Galactic Neutron Star," 
\textit{The Astrophysical Journal Letters}, vol. 934, no. 2, p. L17, 2022. 
doi: 10.3847/2041-8213/ac8007.

\bibitem{Romani_2021}
R.~W.~Romani, D.~Kandel, A.~V.~Filippenko, T.~G.~Brink, and W.~K.~Zheng, 
"PSR J1810+1744: Companion Darkening and a Precise High Neutron Star Mass," 
\textit{The Astrophysical Journal Letters}, vol. 908, no. 2, p. L46, February 2021. 
doi: 10.3847/2041-8213/abe2b4

\bibitem{Demorest_2010}
P.~B.~Demorest, T.~Pennucci, S.~M.~Ransom, M.~S.~E.~Roberts, and J.~W.~T.~Hessels, 
"A two-solar-mass neutron star measured using Shapiro delay," 
\textit{Nature}, vol. 467, no. 7319, pp. 1081--1083, October 2010. 
doi: 10.1038/nature09466.


\bibitem{Antoniadis}
Antoniadis, J., Freire, P. C. C., Wex, N., et al.
\newblock {A Massive Pulsar in a Compact Relativistic Binary}.
\newblock {\em Science}, \textbf{340} (6131), 1233232, 2013.
\newblock \url{https://www.science.org/doi/abs/10.1126/science.1233232}

\bibitem{Cromartie}
Cromartie, H. T., Fonseca, E., Ransom, S. M., et al.
\newblock {Relativistic Shapiro delay measurements of an extremely massive millisecond pulsar}.
\newblock {\em Nature Astronomy}, \textbf{4}(1), 72-76, 2020.
\newblock \url{https://doi.org/10.1038/s41550-019-0880-2}


\bibitem{Hinderer}
Hinderer, T., Lackey, B. D., Lang, R. N., Read, J. S.
\newblock {Tidal deformability of neutron stars with realistic equations of state and their gravitational wave signatures in binary inspiral}.
\newblock {\em Phys. Rev. D}, \textbf{81}(12), 123016, 2010.
\newblock \url{https://link.aps.org/doi/10.1103/PhysRevD.81.123016}




\bibitem{Soma_PRD}
Shriya Soma, Lingxiao Wang, Shuzhe Shi, Horst Stöcker, and Kai Zhou,
\textit{Phys. Rev. D} \textbf{107}, 083028 (2023).
\newblock doi: \href{https://doi.org/10.1103/PhysRevD.107.083028}{10.1103/PhysRevD.107.083028}.






\bibitem{Ozel_2016}
Feryal Ozel et al.,
\textit{The Astrophysical Journal} \textbf{820}, 28 (2016).
\newblock doi: \href{https://doi.org/10.3847/0004-637X/820/1/28}{10.3847/0004-637X/820/1/28}.

\bibitem{rNattila}
J. Nättilä et al.,
\textit{Astronomy \& Astrophysics} \textbf{608}, A31 (2017).
\newblock doi: \href{https://doi.org/10.1051/0004-6361/201731082}{10.1051/0004-6361/201731082}.


\bibitem{Gonzalez}
Denis González-Caniulef et al.,
\textit{Monthly Notices of the Royal Astronomical Society} \textbf{490} (4), 5848-5859 (2019).
\newblock doi: \href{https://doi.org/10.1093/mnras/stz2941}{10.1093/mnras/stz2941}.



\bibitem{Biswas_2022}
Biswas, B.
\newblock{Bayesian Model Selection of Neutron Star Equations of State using Multi-messenger Observations}.
\newblock {\em The Astrophysical Journal}, (926)(1),75,2022
\newblock \url{https://dx.doi.org/10.3847/1538-4357/ac447b}



\bibitem{Lim2019}
Lim, Y. and Holt, J. W.
\newblock{Bayesian modeling of the nuclear equation of state for neutron star tidal deformabilities and GW170817}.
\newblock {\em The European Physical Journal A}, \textbf{55}(11),209,2019
\newblock \url{https://doi.org/10.1140/epja/i2019-12917-9}

\bibitem{Raithel}
Raithel, C. et al.
\newblock{From Neutron Star Observables to the Equation of State. II. Bayesian Inference of Equation of State Pressures}.
\newblock {\em The Astrophysical Journal}, \textbf{844}(2),156,2017


\bibitem{Char}
Char, P., Traversi, S., Pagliara, G.
\newblock {A Bayesian Analysis on Neutron Stars within Relativistic Mean Field Models}.
\newblock {\em Particles}, \textbf{3}(3), 621-629, 2020.
\newblock \url{https://www.mdpi.com/2571-712X/3/3/40}


\bibitem{Goodfellow-et-al-2016}
Goodfellow, I., Bengio, Y., Courville, A.
\newblock {\em Deep Learning}.
\newblock MIT Press, 2016.
\newblock \url{http://www.deeplearningbook.org}


\bibitem{dozat2016}
Dozat, Timothy. "Incorporating Nesterov Momentum into Adam." In *International Conference on Learning Representations (ICLR) Workshop*, 2016.


\bibitem{chollet}
Chollet, F.
\newblock {\em Keras}.
\newblock GitHub, 2015.
\newblock \url{https://github.com/fchollet/keras}



\bibitem{tensorflow}
TensorFlow Authors, "TensorFlow: Large-Scale Machine Learning on Heterogeneous Systems," Software available from tensorflow.org (2015).




\bibitem{Ma_2019}
Yong-Liang Ma and Mannque Rho,
\textit{Phys. Rev. D} \textbf{100}, 114003 (2019).
\newblock doi: \href{https://doi.org/10.1103/PhysRevD.100.114003}{10.1103/PhysRevD.100.114003}.


\bibitem{Ma_2021}
Hyun Kyu Lee, Yong-Liang Ma, Won-Gi Paeng, and Mannque Rho,
\textit{Mod. Phys. Lett. A} \textbf{37}, 2230003 (2022).
\newblock doi: \href{https://doi.org/10.1142/S0217732322300038}{10.1142/S0217732322300038}.


\bibitem{Traversi2022}
S.~Traversi, P.~Char, G.~Pagliara, and A.~Drago, 
"Speed of sound in dense matter and two families of compact stars," 
\textit{Astronomy \& Astrophysics}, vol. 660, p. A62, April 2022. 
doi: 10.1051/0004-6361/202141544.


\bibitem{Ecker_trace}
C.~Ecker and L.~Rezzolla, 
"Impact of large-mass constraints on the properties of neutron stars," 
\textit{Monthly Notices of the Royal Astronomical Society}, vol. 519, no. 2, pp. 2615--2622, February 2023. 
doi: 10.1093/mnras/stac3755.



\bibitem{Ecker_2022}
C. Ecker \& L. Rezzolla, "A General, Scale-independent Description of the Sound Speed in Neutron Stars," Astrophys. J. Lett. 939(2), L35 (2022), doi:10.3847/2041-8213/ac8674.


\bibitem{Annala_2023}
E.~Annala, T.~Gorda, J.~Hirvonen, O.~Komoltsev, A.~Kurkela, J.~Nättilä, and A.~Vuorinen, 
"Strongly interacting matter exhibits deconfined behavior in massive neutron stars," 
\textit{Nature Communications}, vol. 14, article no. 8451, 2023. 
doi: 10.1038/s41467-023-38451-3.


\bibitem{Cai}
B.-J.~Cai, B.-A.~Li, and Z.~Zhang, 
"Central speed of sound, the trace anomaly, and observables of neutron stars from a perturbative analysis of scaled Tolman-Oppenheimer-Volkoff equations," 
\textit{Physical Review D}, vol. 108, no. 10, p. 103041, November 2023. 
doi: 10.1103/PhysRevD.108.103041.



\bibitem{Fujimoto_Trace}
Y. Fujimoto, K. Fukushima, L. D. McLerran, and M. Prasza\l{}owicz,
\textit{Phys. Rev. Lett.} \textbf{129}, 252702 (2022).

\bibitem{Tews2018}
I.~Tews, J.~Carlson, S.~Gandolfi, and S.~Reddy, 
"Constraining the Speed of Sound inside Neutron Stars with Chiral Effective Field Theory Interactions and Observations," 
\textit{The Astrophysical Journal}, vol. 860, no. 2, p. 149, 2018. 
doi: 10.3847/1538-4357/aac267.





\bibitem{Ferrer2023}
E.~J.~Ferrer and A.~Hackebill, 
"Speed of sound for hadronic and quark phases in a magnetic field," 
\textit{Nuclear Physics A}, vol. 1031, p. 122608, March 2023. 
doi: 10.1016/j.nuclphysa.2023.122608.

\bibitem{Bedaque2015}
P.~Bedaque and A.~W.~Steiner, 
"Sound Velocity Bound and Neutron Stars," 
\textit{Physical Review Letters}, vol. 114, p. 031103, 2015. 
doi: 10.1103/PhysRevLett.114.031103.

\bibitem{Reed2020}
B.~Reed and C.~J.~Horowitz, 
"Large sound speed in dense matter and the deformability of neutron stars," 
\textit{Physical Review C}, vol. 101, p. 045803, 2020. 
doi: 10.1103/PhysRevC.101.045803.






















\end{thebibliography}
\end{document}